\documentclass{article}

\usepackage{microtype}
\usepackage{graphicx}
\usepackage{subfigure}
\usepackage{booktabs} 
\usepackage{svg}
\usepackage{rotating}
\usepackage{colortbl}
\usepackage{multirow}
\usepackage{makecell}

\usepackage[acronym]{glossaries}
\usepackage{float}
\usepackage{caption}

\usepackage{hyperref}

\usepackage[accepted]{icml2023}

\usepackage{amsmath}
\usepackage{amssymb}
\usepackage{mathtools}
\usepackage{amsthm}

\usepackage{makecell}

\usepackage{enumitem,xcolor}
\usepackage{tikz}
\usepackage{booktabs}

\usepackage[capitalize,noabbrev]{cleveref}

\newcommand\extrafootertext[1]{%
    \bgroup
    \renewcommand\thefootnote{\fnsymbol{footnote}}%
    \renewcommand\thempfootnote{\fnsymbol{mpfootnote}}%
    \footnotetext[0]{#1}%
    \egroup
}

\theoremstyle{plain}

\theoremstyle{definition}

\theoremstyle{remark}

\newcommand{\seqlength}{T}
\newcommand{\numlayers}{L}
\newcommand{\layerindex}{\ell}
\newcommand{\mask}{M}
\newcommand{\dependencyfunc}{D}
\newcommand{\pdf}{F}
\newcommand{\bucketcost}{q}
\newcommand{\backlog}{b}
\newcommand{\blockspersec}{\mu}
\newcommand{\framerate}{\rho}
\newcommand{\scheduler}{\mathcal{S}}
\newcommand{\forwardspan}{K}
\newcommand{\hiddenvector}{h}
\newcommand{\tempparam}{\tau}
\newcommand{\centers}{o}
\newcommand{\computenode}{v}
\newcommand{\regularizationconst}{\lambda}

\newacronym{architecturename}{ANCAT}{Adaptive Non-Causal Attention Transducer}

\usepackage[textsize=tiny]{todonotes}

\icmltitlerunning{Adaptive Non-causal Transformers for Streaming Neural Transducers}

\begin{document}

\twocolumn[
\icmltitle{Lookahead When It Matters: Adaptive Non-causal Transformers\\for Streaming Neural Transducers}



\icmlsetsymbol{equal}{*}

\begin{icmlauthorlist}
\icmlauthor{Grant P. Strimel}{equal,amazon}
\icmlauthor{Yi Xie}{equal,amazon}
\icmlauthor{Brian King}{equal,amazon}
\icmlauthor{Martin Radfar}{amazon}
\icmlauthor{Ariya Rastrow}{amazon}
\icmlauthor{Athanasios Mouchtaris}{amazon}
\end{icmlauthorlist}

\icmlaffiliation{amazon}{Amazon Alexa, USA}

\icmlcorrespondingauthor{Grant P. Strimel}{gsstrime@amazon.com}
\icmlcorrespondingauthor{Yi Xie}{yixiey@amazon.com}

\icmlkeywords{Causal and Non-causal Attention, Transformers, Speech Recognition, Adaptive Compute}

\vskip 0.3in
]



\printAffiliationsAndNotice{\icmlEqualContribution} 

\begin{abstract}
Streaming speech recognition architectures are employed for low-latency, real-time applications. Such architectures are often characterized by their causality.
Causal architectures emit tokens at each frame, relying only on current and past signal, while non-causal models are exposed to a window of future frames at each step to increase predictive accuracy.
This dichotomy amounts to a trade-off for real-time Automatic Speech Recognition (ASR) system design: profit from the low-latency benefit of strictly-causal architectures while accepting predictive performance limitations, or realize the modeling benefits of future-context models accompanied by their higher latency penalty.
In this work, we relax the constraints of this choice and present the \acrfull{architecturename}.
Our architecture is non-causal in the traditional sense, but executes in a low-latency, streaming manner by dynamically choosing when to rely on future context and to what degree within the audio stream.
The resulting mechanism, when coupled with our novel regularization algorithms, delivers comparable accuracy to non-causal configurations while improving significantly upon latency, closing the gap with their causal counterparts.
We showcase our design experimentally by reporting comparative ASR task results with measures of accuracy and latency on both publicly accessible and production-scale, voice-assistant datasets.
\end{abstract}

\section{Introduction}
\label{sec:introduction}
Modern speech recognition applications have leveraged the benefits afforded by fully-neural architectures to drive enhanced user experiences.
For example, these architectures allow popular virtual assistants such as Amazon Alexa, Google Assistant, and Apple Siri to rely on robust, far-field speech recognition as their primary medium of real-time interaction.
Similarly, offline services such as recorded call transcription and video caption generation apply these architectures as well.
The dominant techniques like Neural-Transducers, Listen-Attend-Spell, and Transformers are all principally sequence-to-sequence models consisting of an audio encoder mapping acoustic input to high-level representations for autoregressive decoding~\cite{Graves2012,Chan2016,Mohamed2019,Chorowski2015,Han2020}.
And while each neural ASR architecture has its individual design elements, generally speaking, each is categorized or implemented as a causal or non-causal model.

Causal ASR models are streamable designs which emit predictions for each frame as they arrive from the audio signal, without access to future frames~\cite{He2018,Yu2021,Radfar2022}.
These models are referred to as causal in the sense that each frame prediction is only dependent on its left (past) context.
Causal configurations are particularly attractive for real-time processing applications where fast response times are essential for the user experience, such as virtual assistants or live stream caption generation.
Non-causal models, on the other hand, have access to information from future frames.
Non-causal models can be either streaming with a bounded number of lookahead frames or full-context with access to the entirety of the audio signal, before beginning to decode~\cite{Zhang2020,Moritz2020,Yeh2019,Tripathi2020}.
Naturally, these models can deliver significantly improved accuracy over their causal counterparts by leveraging future information to disambiguate predictions holistically.
The accuracy gains of non-causal processing typically are accompanied by a steep price of higher latency in real-time applications however, which can hinder these models for production deployment.

In this paper, we introduce the \acrfull{architecturename}, a neural ASR architecture which has the accuracy improvements witnessed by non-causal models, while yielding a latency that is more comparable to streaming causal architectures.
\acrshort{architecturename} achieves this improvement by providing flexibility to the model to rely on variable future context at each frame. 
However, it is trained to be selective about how and when it does so, taking into account both the accuracy and latency implications to poll for future context only where necessary to make accurate predictions.
The resulting behavior is that the model's latency costs for adaptively ingesting lookahead context, when aggregated over the entire frame sequence, is comparatively minimal.
Our model is trained fully end-to-end, jointly learning how and where to apply non-causal context in tandem with its traditional token prediction objectives.

After summarizing motivating related work in Section~\ref{relatedwork}, the remainder of the paper is organized according to our contributions:
Section~\ref{design} introduces the \acrshort{architecturename} design and key architectural concepts,
Section~\ref{lossfuncs} derives novel loss functions developed for training our architecture and regularizing future context in connection with different notions of latency commonly applied in the literature, 
and Section~\ref{experiments} presents empirical results which justify our design elements and showcase the modeling capabilities of \acrshort{architecturename} for streaming ASR tasks on both open-source and industrial data.

\section{Related Work}
\label{relatedwork}
\textbf{Causal, Non-causal, and Streamable ASR.} 
Bridging the accuracy benefits of non-causality with the low-latency deployability of causal models has been the focus of many prior studies.
Several works~\cite{Audhkhasi2021, Moritz2021, Yu2021a} find that distillation techniques leveraging non-causal right context benefit the training of fully causal models.
For instance, ``dual-mode'' ASR, where a single end-to-end model is jointly trained with full-context mode using shared weights, improves causal ASR accuracy~\cite{Yu2021a}.

Other approaches enable more accurate streaming by permitting ingestion of a finite amount of frame-wise lookahead context at the cost of a latency penalty.
For example, using a fixed right-context window of lookahead frames at each layer is common~\cite{Zhang2020}.
Chunking approaches are also effective for non-causality~\cite{Tsunoo2019, Dong2019, Shi2021, Chen2021}.
With chunking, the stream is broken down into fixed-sized groupings of non-overlapping, adjacent frames.
Within a chunk, all frames have access to one another and possibly frames from prior chunks.
Similar to dual-mode training,~\cite{Swietojanski2022} also extends in-place distillation to chunking of variable sizes, training a single model but allowing different chunk sizes to be configured at inference to match hardware specifications.
Scout-networks~\cite{Wang2020}, meanwhile, use word boundaries to define the position and sizes of chunks and use a separate network trained on forced alignment data to predict 
boundaries at inference time.

Another set of approaches unify causal and non-causal context into one system by training streaming and full-context encoders and stacking them.~\cite{Narayanan2021} and~\cite{Li2021} propose ``cascaded'' audio encoders where a causal first-pass encoder is followed by a stacked non-causal encoder operating on the first encoder's outputs.
These modules are jointly learned to produce accuracy improvements, but accept the latency impact from the full-context second-pass encoder.

\textbf{Adaptive Neural Compute.}
Adaptive compute, also referred to commonly as variable or dynamic compute, is a technique that adjusts the amount of neural computation a model executes during inference as a function of each individual input.
The motivating intuition of the approach is that since each instance has different characteristics, the corresponding amount of computation expended should reflect this variety, conditioning for more resources and operations only where necessary.
Researchers have explored these ideas extensively across machine learning areas, including 
NLP~\cite{Graves2016, Jernite2017, Dehghani2019, Elbayad2020},
vision~\cite{Bolukbasi2017, Figurnov2017}, speech~\cite{Macoskey2021a, Macoskey2021,Xie2022,Peng2023}
and recommendation systems~\cite{Song2019}.
We refer readers to~\cite{Han2021} for a comprehensive survey.
While \cite{Sukhbaatar2020, Chang2020} also propose learning static attention span adjustment, our mechanism will vary the span across inputs and frames adaptively for streaming.
Additionally, all adaptive techniques effectively tie their compute to a particular cost metric, such as floating point operations.
For our application, we will show how to link our dynamic compute mechanism to key latency measures for real-time speech recognition.  


\textbf{Latency Measures.} 
There are intricacies to assessing the latency of a streaming ASR architecture, and as a result, numerous  metrics have been proposed to encapsulate its various facets.
User-perceived latency (UPL), endpointer latency, first token emission delay, algorithmic latency, mean alignment delay, and partial recognition latency are all among those measures which are considered in the literature~\cite{Shangguan2021, Sainath2020, Inaguma2020, Yu2021}.
While each definition attempts to capture a different latency driver,
for models using non-causal context, it is natural to use algorithmic latency, and therefore, we also adopt this measure to inform the design of our adaptive non-causal streaming architecture. 
\textit{Algorithmic latency} reports the time required processing the input to produce the output of an audio frame -- frame length combined with the amount of lookahead frames used~\cite{Shi2021}.
Since our algorithm is nondeterministic, we adjust to reporting a statistical version of the metric.
We also consider a \textit{compute-induced UPL} metric in our design, detailed in Section~\ref{subsec:computeuplloss}, which jointly accounts for processing speed and additional work scheduled by our non-causal mechanism to derive their combined impact on UPL.

\section{Adaptive Non-causal Design}
\label{design}
In this work we focus on the Neural Transducer architecture consisting classically of three modules: an encoder network, a prediction network, and a joint network.
The encoder network takes a sequence of $\seqlength$ feature frame vectors $\left(x_1, \dots, x_\seqlength \right)$ extracted from the audio signal and maps it to a corresponding sequence of high-level acoustic representations.
The prediction network operates autoregressively over a sequence of $U$ labels  $\left(y_1, \dots , y_U \right)$.
The joint network combines the outputs from the encoder and prediction networks to model the likelihood for the next label emitted.

The transducer architecture is a popular choice for real-time applications because the encoder network can be trained for streaming, allowing the joint network to emit its prediction on the $i$-th frame relying on a bounded number of future frames; however, full-context encoders can be applied to yield additional accuracy for non-streaming scenarios.
Our design utilizes Transformer-based encoders which are constructed from $\numlayers$ stacked blocks (a.k.a. layers, terms which we use interchangeably), each of which accepts the output from the previous block and produces hidden vector $\hiddenvector_i^\layerindex$ for frame $i$ and layer $\layerindex$.
Transformer-based blocks consist of various compositions of sublayers such as feed-forwards, layer-norms, and even convolutions in the case of Conformer~\cite{Gulati2020}, but all contain the multi-headed self-attention (MHSA) sublayer.

\subsection{Compute DAGs and Attention Masking}
Transformer-based attention architectures define a natural compute graph of input dependencies for each output of each layer.
Under this directed acyclic graph (DAG) representation, each vertex $\computenode_i^\layerindex$ (or node) represents the compute on frame $i$ of layer $\layerindex$, and a directed edge $\left(\computenode_j^{\layerindex-1}, \computenode_i^\layerindex\right)$ between vertices of adjacent layers indicates the reliance on the output of $\computenode_j^{\layerindex-1}$ for computing $\computenode_i^\layerindex$, namely that $\hiddenvector_j^{\layerindex - 1}$ (or a mapping thereof) is attended over when computing $\hiddenvector_i^{\layerindex}$.

Our \acrshort{architecturename} architecture is designed to dynamically and strategically fill in the edges of this compute DAG, balancing both accuracy and latency considerations.
To accomplish this, we train the traditional encoder weights while introducing jointly trainable schedulers $\scheduler^\layerindex$ into the architecture, one for each layer $\layerindex$.
Each scheduler $\scheduler^\layerindex$ accepts the prior layer's result $\hiddenvector_i^{\layerindex-1}$ and determines the future, non-causal inputs to attend over to compute $h_i^\layerindex$.
Hence, $\scheduler^\layerindex(h_i^{\layerindex-1})$ predicts the non-causal edges from vertices $\computenode_{j}^{\layerindex - 1}$ to vertex $\computenode_i^{\layerindex}$ for positions $j > i$.

\begin{figure}[ht]
  \centering
  \includegraphics[width=0.95\columnwidth]{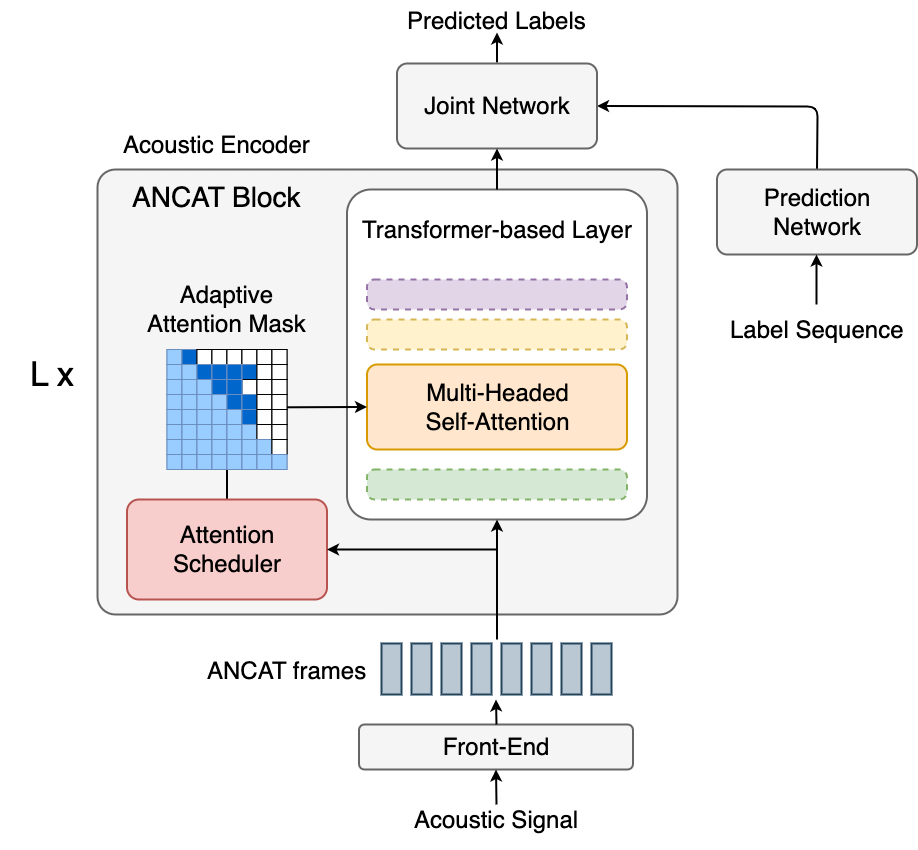}
  \caption{\acrfull{architecturename}. The architecture is a neural transducer with an acoustic encoder of $\numlayers$ stacked Transformer-based blocks where each layer is augmented with an attention scheduler. Each scheduler learns to fill in the right context attention connections and passes the resulting mask to the multi-headed self-attention.}
\end{figure}

During training, we leverage a series of attention masks $\mask^\layerindex \in \left[0, 1\right]^{\seqlength \times \seqlength}$ to represent these edges, where $\mask^\layerindex_{i,j} = 1$ indicates the full presence of the $(\computenode_j^{\layerindex-1}, \computenode_i^\layerindex)$ edge and $0$ indicates its absence:
\begin{align*}
\mask^\layerindex_{i,j} =
\begin{cases}
    1& j \leq i\\
    \scheduler^\layerindex(h_i^{\layerindex-1})_j & \text{otherwise}
\end{cases}
\quad 1 \leq i, j \leq \seqlength
\end{align*}

The mask is applied for computing attention scores across each of the attention heads in the MHSA sublayer. 
Namely, for a scaled dot product value matrix $P$ computed from query matrix $X_{\text{query}}$ and key matrix $X_{\text{key}}$ for a particular head
\begin{align*}
P = \frac{X_{\text{query}} X_{\text{key}}^\top}{\sqrt{d}},
\end{align*}
the masked attention scores can be computed as
\begin{align*}
A_{i,j} = \frac{ \exp{\left( P_{i,j} \right)} \mask_{i,j}^\layerindex}{\sum_{t}^\seqlength \exp{\left( P_{i,t} \right)} \mask_{i,t}^\layerindex} \quad 1 \leq i, j \leq \seqlength
\end{align*}
for each attention head of layer $\layerindex$ or using method of~\cite{Xie2022}.

\subsection{Learned Schedulers}
The role of the scheduler is to fill in the forward (right of the main diagonal) values of its corresponding attention mask.
To simplify the learning process, instead of predicting each connection individually, we view each scheduler $\scheduler$ as estimating the length of the non-causal attention span to apply at its layer (i.e., the number of future frames to consider) on a particular input.
We employ a smooth, differentiable, reverse ``S''-shaped function decreasing from $1$ to $0$ with an adjustable parameter $\tempparam$ to control the sharpness of the curve.
The schedulers learn where to shift the center $\centers$ of this curve over a span of $\forwardspan$ maximum lookahead frames.
Specifically, $\scheduler^\layerindex$ is computed as 
\begin{align*}
    \centers_i^\layerindex &=\sigma\left(\text{FFN$^\layerindex$}(h_i^{\layerindex-1})\right) \left(\forwardspan + \epsilon\right) \\
    \scheduler^\layerindex(h_i^\layerindex)_j &= 
    \begin{cases}
    1 - \sigma\left(\left(j - i - \centers_i^\layerindex\right) / \tempparam \right) &j \leq \forwardspan\\
    0 & \text{otherwise,}
    \end{cases}
\end{align*}

where $\sigma$ is the standard sigmoid function and $\text{FFN}^\layerindex$ is a learnable feed-forward network consisting of two linear transforms with a non-linear activation in between, with the second transform projecting to a single scalar.
The small constant $\epsilon$ is used for enabling the $\forwardspan$th lookahead frame to potentially have a near-full connection while maintaining numeric stability during training.

Note that the design permits ``soft edges'' during the learning process with mask values between $0$ and $1$.
However, $\tempparam$ can be annealed towards $0$ to gradually morph all soft edges into definitive binary ones.
Our loss function design will leverage this property as well.

\section{Regularizing Future Context}
\label{lossfuncs}
Complementing the design of the \acrshort{architecturename} architecture, we craft several loss functions which regularize the decisions of the schedulers to explicitly account for different notions of latency for streaming systems.

\subsection{Naive Regularization}
\label{l1_train}
To balance both the primary training objective of the neural transducer with the incurred impact of the schedulers' ingestion of non-causal frames, one can modify the training loss function to the form
\begin{align*}
\mathcal{L} = \mathcal{L}_{\text{transducer}} + \regularizationconst \mathcal{L}_{\text{sched}}
\end{align*}
A first attempt would be to simply regularize over all lookahead connections, with the intuition that attending over the fewest possible future frames across layers will yield lower latency
\[
\mathcal{L}_\text{L1} = \frac{1}{T}\sum_{\layerindex} \sum_{j > i} \mask^\layerindex_{i,j}
\]
This L1-type regularization will serve as a baseline for our experiments.
However, indiscriminately regularizing in this manner does not account for a connection's non-local effects on latency.
For example, a forward edge present in a lower level can dramatically impact the prediction delay over many frames because of the propagating effects layer over layer.
Furthermore, the simple regularization above is not directly tied to an explicit latency measure.

\begin{figure}[t]
\includegraphics[width=0.95\columnwidth]{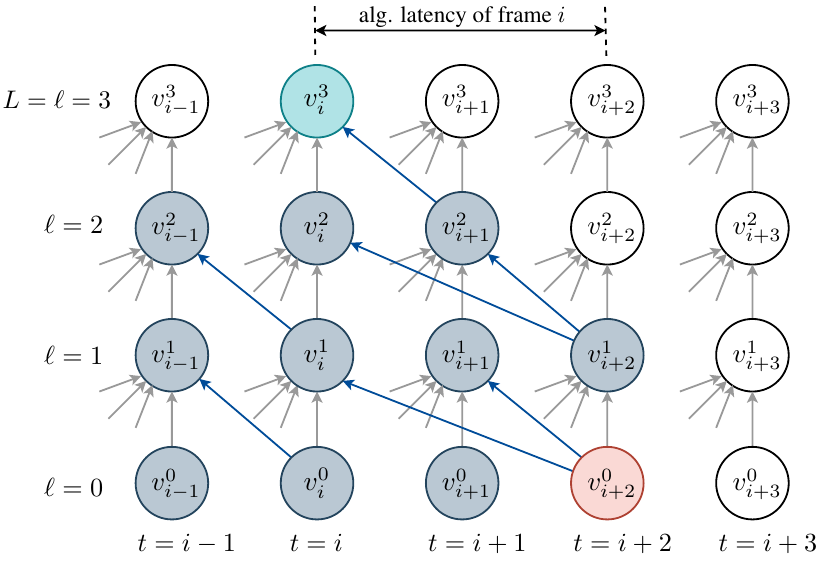}
  \caption{Compute graph example of \acrshort{architecturename} for three layers. Non-causal edges (blue) are provided by the learned schedulers. Dependency nodes (gray) of $v_i^L$ (cyan) determine the algorithmic latency. The difference $\delta(i)$ between the final dependency frame (red) and $v_i^L$ determines the algorithmic latency for frame $i$.}
\end{figure}

\subsection{Algorithmic Latency}
\label{subsec:alglat}
Here, we formulate how to regularize with respect to the notion of algorithmic latency.
Recall that traditionally, algorithmic latency would refer to the minimal theoretic amount of time the model needs to wait 
before it can emit a symbol at a particular frame, resulting from the number of non-causal future frames relied upon $\delta(i)$ for prediction at each timestep $i$ along with the frame length (which is a constant).\footnote{It is common to include the ``current'' $i$-th frame in this calculation; however, for  clarity of presentation under our setup, we omit it. But since this singular frame amounts to an additive constant, one can easily translate any result to account for it.}
This delay is a deterministic value for standard architectures.
In contrast, with our \acrshort{architecturename} architecture, the algorithmic latency can vary from frame to frame, and we therefore adopt a time-averaged mean definition:
\begin{equation*}
\text{algorithmic latency} = \frac{1}{\seqlength} \sum_{i} (\text{frame length}) \delta(i)
\end{equation*}
We derive an algorithm below to directly integrate the algorithmic latency stemming from the schedulers' decisions and aggregate the result into a regularization loss we apply during training.

To do so, we first define the notion of a dependency function.
We define $\dependencyfunc^{\layerindex}_{i,j} \in [0,1]$ to represent the (fractional) dependency for computing $\hiddenvector_{i}^\layerindex$ on the input node $v_{j}^0$.
In other words, $\dependencyfunc^{\layerindex}_{i,j}$ represents if $\computenode_{j}^0$ is connected to $\computenode_{i}^\layerindex$ by some pathway through the compute DAG.
The dependency values can be fractional because the edges will be fractional during training.

For the \acrshort{architecturename} architecture, we propose the following memoized dynamic programming formulation using the attention masks to recursively compute the dependency matrices at each layer
\begin{align*}
    \dependencyfunc^{\layerindex}_{i,j} = 
\begin{cases}
    \mask^{\layerindex}_{i,j}& \layerindex = 1\\
    \max\limits_{t}\enskip{\mask^{\layerindex}_{i,t} \cdot \dependencyfunc^{\layerindex - 1}_{t,j}} & \layerindex > 1
\end{cases}
\end{align*}

This algorithm can be viewed as computing the maximum fractional strength of a pathway between an input frame and a compute node.
Interestingly, this algorithm becomes a special case reduction of a classic shortest path in a graph problem using edge weight from $\computenode_j^{\layerindex-1}$ and $\computenode_i^\layerindex$ as $-\ln{\mask^{\layerindex}_{i,j}}$ and with the final $\dependencyfunc^{\layerindex}_{i,j}$ values extracted as $\exp\left\{- \text{distance}\left(\computenode_j^{0}, \computenode_i^\layerindex\right)\right\}$.

Importantly, for our application, the algorithm produces dependency matrices for each layer $\layerindex$ that has $\dependencyfunc^{\layerindex}_{i,j}$ as monotonically decreasing in $j$.
The following argument by induction proves this fact:
The base case is straight forward because $\dependencyfunc^{1}_{i,j} = \mask^{1}_{i,j}$ and by scheduler function monotonicity. Now let $\dependencyfunc^{\layerindex}_{i,j} = \mask^{\layerindex}_{i,t} \cdot \dependencyfunc^{\layerindex - 1}_{t,j}$ for some $t$. Since $\dependencyfunc^{\layerindex - 1}_{t,j-1} \geq \dependencyfunc^{\layerindex - 1}_{t,j}$ by induction hypothesis, there exists at least one $t'$, namely $t'=t$, over which a maximum is taken such that $\dependencyfunc^{\layerindex}_{i,j-1} \geq \mask^{\layerindex}_{i,t'} \cdot \dependencyfunc^{\layerindex - 1}_{t',j-1} \geq \mask^{\layerindex}_{i,t} \cdot \dependencyfunc^{\layerindex - 1}_{t,j} = \dependencyfunc^{\layerindex}_{i,j}$.

As a result of this monotonicity, we treat $\dependencyfunc^{\layerindex}_{i,j}$ as an approximate, inverse cumulative distribution function over $j$ where $\hat{P}(\text{last dependency frame of $\computenode_{i}^\layerindex$} \leq t) = 1 - \dependencyfunc^{\layerindex}_{i,t}$.
Therefore, the corresponding density function $\pdf^{\layerindex}_{i,j} = \dependencyfunc^{\layerindex}_{i,j-1} - \dependencyfunc^{\layerindex}_{i,j}$ provides a natural weighting for the position of the final lookahead frame required to compute node $\computenode^{\layerindex}_{i}$.

Observe that indeed $\sum_j \pdf^{\layerindex}_{i,j} = 1$ for all $i, \layerindex$ and that the distribution turns into a strictly one-hot vector coding which indicates the exact position of the last frame dependency of $\computenode^{\layerindex}_{i}$
as $\tempparam$ is annealed to produce binary connections from the schedulers.

We now can use this algorithm to directly regularize mean (fractional) algorithmic latency over all frames of the utterance
\[
\mathcal{L}_{\text{Alg.Lat.}} = \frac{1}{\seqlength} \sum_{i} \hat{\delta}(i) = \frac{1}{\seqlength} \sum_{j > i} \left(j - i \right) \pdf^{\numlayers}_{i,j}
\]
using $\pdf^{\numlayers}_{i,:}$ to indicate (weight) the last frame dependencies of the top layer $\numlayers$ at each frame.

\subsection{Compute-Induced UPL}
\label{subsec:computeuplloss}
While algorithmic latency as a metric provides perspective on the minimum delay a streaming ASR model is to expect, it operates under the assumption that all compute is instantaneous, and therefore, only serves as a lower bound of the model's response UPL. 
To more realistically model and regularize with respect to UPL, we propose a loss which also accounts for processing speed.

We denote $\blockspersec$ as the effective processor speed in terms of compute nodes (encoder layers) which can be processed each second and $\framerate$ as the feature frame rate in frames per second.
We use these constants and the algorithm presented in Section~\ref{subsec:alglat} to compute at which timestep each node becomes available to establish an accounting of the compute backlog.
The final amount of work remaining (nodes left to compute) in the backlog on the final frame will then be used to regularize the UPL.

\begin{figure}[htbp]
  \centering
 \includegraphics[width=1\columnwidth]{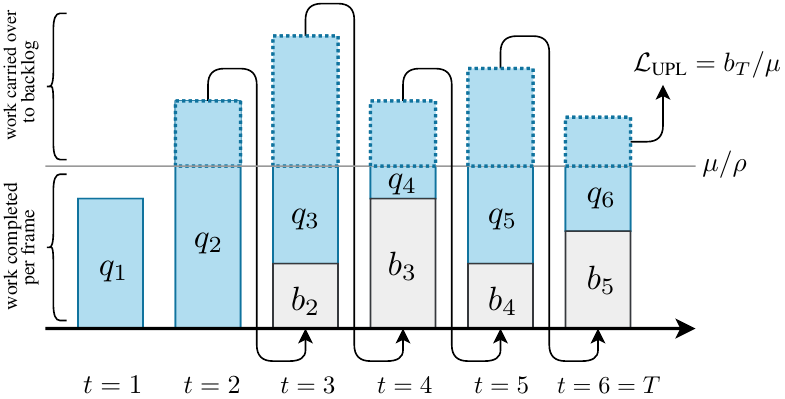}
  \caption{Example of the UPL algorithm in action. Work which cannot be completed on the current frame under a specified throughput is carried over to the subsequent frame. The remaining work at the end of the utterance defines the UPL.}
  \label{fig:backlog}
\end{figure}

Letting $\bucketcost_i$ to be the number of nodes which have their input dependencies met on frame $i$, we have
\begin{align*}
\bucketcost_i = \sum_{\ell} \sum_{t < i} \pdf^{\ell}_{t,i}
\end{align*}
which is the amount of new work to be added to the backlog at $i$. Defining $\backlog_i$ to be the buffered node backlog (lag accumulated at the $i$-th timestep), using a method similar to that of \cite{Macoskey2021}, we express the compute-induced UPL loss as the response delay in seconds $\mathcal{L}_\text{UPL} = {\backlog_\seqlength}/{\mu}$
with $\backlog_i$ derived recursively as
\begin{align*}
    b_i = 
\begin{cases}
    0& i = 0\\
    \max \enskip \left\{\backlog_{i - 1} + \bucketcost_i - \blockspersec/\framerate, 0 \right\} & i > 0
\end{cases}
\end{align*}
where $\blockspersec/\framerate$ is the compute throughput in terms of how many nodes per timestep can be burned down in the backlog.
Figure \ref{fig:backlog} illustrates how this UPL computation is carried out.

\begin{table*}[ht!]
\caption{Results on LibriSpeech test ``clean'' for Conformer and T-T encoder backbones.
The models are trained with different streaming settings for maximum lookahead frames $\forwardspan$ on each layer, while the full context model is also reported.
\textit{Layerwise} and \textit{Chunked} represent standard lookahead attention structures and chunked-aware attention methods of prior literature.
\acrshort{architecturename} models trained with L1 regularization and our novel loss are shown as \acrshort{architecturename}\textit{-L1} and \acrshort{architecturename}\textit{-Alg.Lat.}, respectively. 
\acrshort{architecturename}\textit{-Alg.Lat.} improves WER by significant margins over nearest competitors with comparable mean algorithmic latency and likewise for 50th and 90th percentiles.}
\centering
\scalebox{0.95}{
\begin{tabular}{c|p{0.5cm}p{0.5cm}p{0.5cm}p{0.5cm}p{0.5cm}p{0.5cm}p{0.5cm}p{0.5cm}p{0.5cm}p{0.5cm}p{0.5cm}p{0.5cm}p{0.5cm}p{0.5cm}}
\specialrule{.2em}{.1em}{.1em} 
\rule{0pt}{15pt}  & \multicolumn{14}{c}{Conformer}                   \\ 

\hline

\rule{0pt}{15pt} $\forwardspan$ ($\#$ lookahead) &\multicolumn{1}{c}{0}&\multicolumn{4}{c}{2}&     \multicolumn{4}{c}{5}&\multicolumn{4}{c}{10}  &\multicolumn{1}{c}{Full}           \\ 
\cmidrule(r{2pt}){1-1}\cmidrule(l{2pt}){2-2}\cmidrule(l){3-6}\cmidrule(l){7-10}\cmidrule(lr){11-14}\cmidrule(){15-15}\\

Model         &  
\multicolumn{1}{c}{\rotatebox[origin=l]{90}{\tikz\draw[color=red!80, fill=red!40] (0,0) circle (.5ex); \textit{Causal}\strut} }&  
\multicolumn{1}{c} {\rotatebox[origin=l]{90}{\tikz\draw[magenta!80,fill=magenta!20] (0,-1) circle (.5ex); \textit{Layerwise}\strut}}   & 
\multicolumn{1}{c} {\rotatebox[origin=l]{90}{\tikz\draw[magenta!80,fill=magenta!20] (0,0) circle (.5ex); \textit{Chunked}\strut}} & 
\multicolumn{1}{c} {\rotatebox[origin=l]{90}{\tikz\draw[magenta!80,fill=magenta!20] (0,0) circle (.5ex); \acrshort{architecturename}-\textit{L1}\strut}} &  
\multicolumn{1}{c} {\rotatebox[origin=l]{90}{\tikz\draw[magenta!80,fill=magenta!20] (0,0) circle (.5ex); \acrshort{architecturename}-\textit{Alg.Lat.}\strut}} &  
\multicolumn{1}{c} {\rotatebox[origin=l]{90}{\tikz\draw[color=orange!80, fill=orange!20] (0,0) circle (.5ex); \textit{Layerwise}\strut}} & 
 \multicolumn{1}{c} {\rotatebox[origin=l]{90}{\tikz\draw[color=orange!80, fill=orange!20] (0,0) circle (.5ex); \textit{Chunked}\strut}} & 
 \multicolumn{1}{c} {\rotatebox[origin=l]{90}{\tikz\draw[color=orange!80, fill=orange!20] (0,0) circle (.5ex); \acrshort{architecturename}-\textit{L1}\strut}} & 
 \multicolumn{1}{c} {\rotatebox[origin=l]{90}{\tikz\draw[color=orange!80, fill=orange!20] (0,0) circle (.5ex); \acrshort{architecturename}-\textit{Alg.Lat.}\strut}}& 
 \multicolumn{1}{c}{\rotatebox[origin=l]{90}{\tikz\draw[color=blue!80, fill=blue!20] (0,0) circle (.5ex);  \textit{Layerwise}\strut} }& 
 \multicolumn{1}{c}{\rotatebox[origin=l]{90}{\tikz\draw[color=blue!80, fill=blue!20] (0,0) circle (.5ex); \textit{Chunked}\strut} }& 
 \multicolumn{1}{c}{\rotatebox[origin=l]{90}{\tikz\draw[color=blue!80, fill=blue!20] (0,0) circle (.5ex); \acrshort{architecturename}-\textit{L1}\strut} }& 
\multicolumn{1}{c} {\rotatebox[origin=l]{90}{\tikz\draw[color=blue!80, fill=blue!20] (0,0) circle (.5ex); \acrshort{architecturename}-\textit{Alg.Lat.}\strut} }&

\multicolumn{1}{c}{\rotatebox[origin=l]{90}{\tikz\draw[color=cyan!80, fill=cyan!20] (0,0) circle (.5ex); \textit{Full Context}\strut} }\\ 
\cmidrule(r{2pt}){1-1}\cmidrule(l{2pt}){2-2}\cmidrule(l){3-6}\cmidrule(l){7-10}\cmidrule(lr){11-14}\cmidrule(){15-15}\\
WER (\%)          & 6.66  & 4.58   & 6.27 & 5.68  & \textbf{5.31} &  4.23 & 5.68 & 5.43 & \textbf{5.13} & 4.12 & 5.15 & 5.13 & \textbf{4.76}  & 4.09 \\
Alg.Latency (ms)  & 0.00 & 2240 & 63.2  & 60.5 & 60.1 & 3240 & 235 & 235 & 202 & 3520 & 518 & 495 & 498 &  -  \\

Alg.Latency@50 (ms) & 0.00 & 2380 &59.9  & 61.6 & 58.8  & 2940& 236 & 223 & 237  & 2940 & 518 & 488 &  489& -   \\
Alg.Latency@90 (ms) & 0.00 & 2920 &  62.4 & 86.4 & 76.7 &  5680 & 241 & 310 & 274 & 6650 &  541 & 705 & 623 & -  \\ 
$\ell_1$-norm (frames)  & 0.00 & 27.1 & 6.94 & 1.78 & 8.12 & 65.3 & 27.3 & 7.04 & 16.5 & 122 & 60.3 & 16.2 & 59.1 & -  \\\hline \hline
\rule{0pt}{15pt}  & \multicolumn{14}{c}{Transformer}                   \\ 
\cmidrule(r{2pt}){1-1}\cmidrule(l{2pt}){2-2}\cmidrule(l){3-6}\cmidrule(l){7-10}\cmidrule(lr){11-14}\cmidrule(){15-15}\\
WER  (\%)         & 6.90  & 4.88    & 6.56 & 5.92 & \textbf{5.57} & 4.63& 5.74 & 5.67 & \textbf{5.36} & 4.42 & 5.65 & 5.27 & \textbf{5.11}  & 4.55 \\
Alg.latency (ms)  &0.00 & 2240 & 63.2  & 61.2 & 60.5  & 3240 & 235 & 232  & 229 & 3520 & 518 & 512 & 496 &  -  \\
Alg.Latency@50 (ms)        & 0.00 & 2380 & 59.9  & 60.4 & 55.8 & 2940& 236 & 230 &  228& 2940 & 518  & 508  & 484 &  -  \\
Alg.Latency@90 (ms)          & 0.00 & 2920 & 62.4 & 88.6 & 80.2 & 5680 & 241 & 269 & 305 &  6650 &  541 & 701 & 613 & -\\ 
$\ell_1$-norm (frames)           & 0.00 & 27.1& 6.94 &  1.89 & 7.93  & 65.3 & 27.3 & 5.47 & 21.5 & 122 & 60.3 &  18.5 & 36.3 &  -  \\

\end{tabular}}
\label{table:alg_main}
\vspace{-0.25cm}
\end{table*}
\section{Experimental Results}
\label{experiments}
\vspace{-0.15cm}
We organize our results in this section to demonstrate \acrshort{architecturename}'s combined accuracy and algorithmic latency improvements on publicly available ASR data, supplemented with a summary of findings on industry data, model dynamics, and compute-induced UPL studies while referring the reader to corresponding appendices for their full results.
\vspace{-0.1cm}
\subsection{LibriSpeech Experimental Setup}
We investigate our architectures using the LibriSpeech corpus~\cite{Panayotov2015} comprised of 960 hours of training data collected from read audio books.
Our evaluations report on the associated ``clean'' test data.
Audio clips are preprocessed with a 64-dimensional log-filterbank energy feature extractor, and these feature vectors are stacked with a stride size of 2 and downsampled by 3 before being processed by a small convolution front-end to produce 120ms frames as input for the transformer-based blocks.

We conduct our experiments on two popular transformer-based architectures: Conformer~\cite{Gulati2020} and Transformer-Transducer (T-T)~\cite{Zhang2020}.
For both Conformer and T-T, we use encoders consisting of 14 stacked blocks, one-layer 640-unit LSTM prediction networks, and a 512-unit feed-forward joint network.
For our \acrshort{architecturename} variants, we augment each block with learnable schedulers of 64-dimension hidden units. 
The detailed model configurations are listed in Appendix~\ref{append:libri-setting}.

For LibriSpeech, all the models are trained for 150 epochs of 1k steps with the Transformer-based encoder backbone and schedulers trained end-to-end jointly using the Adam optimizer~\cite{Kingma2015} using the same hyperparameters settings specified by~\cite{Gulati2020}.
During training, we also apply SpecAugment~\cite{Park2019} with mask parameter $(\textit{F} = 27)$ and 20 time masks with maximum time mask ratio $(p_S = 0.04)$, where the maximum-size of the time mask is set to $p_S$ times the length of the utterance.
We use a vocabulary of 4097 word pieces and the standard RNN-T beam search decoding with a width of 16.
Further specifications on model configurations and training hyperparameters are shared in Appendix~\ref{append:libri-setting}.

\subsection{Baselines}
We establish baseline models for Conformer and T-T models using causal and full-context attention.
We also build baselines for non-causal streaming mechanisms proposed in prior studies.
These include streaming attention which ingests $\forwardspan$ lookahead frames at each layer~\cite{Zhang2020}, denoted here as \textit{Layerwise}, and chunking transformer attention~\cite{Shi2021,Chen2021}, denoted as \textit{Chunked}, which is a popular and efficient approach for comparison.
Additionally, we present  \acrshort{architecturename} models with basic L1-type regularization over the amount of future frames as described in Section~\ref{l1_train}, marked as \acrshort{architecturename}\textit{-L1}.

\subsection{Algorithmic Latency Results}
Our main empirical result showcases the significant improvements on each operating point of the latency-accuracy trade-off curve that \acrshort{architecturename} produces when trained with our novel regularization loss, and thereby, shifting the Pareto frontier over existing methods.
Namely, for each fixed accuracy target, our \acrshort{architecturename} model produces significant decreases in algorithmic latency over the nearest baseline, and conversely, for a desired algorithmic latency, \acrshort{architecturename} also yields significant improvements in accuracy.
We establish this result by training the baseline and \acrshort{architecturename} models under various hyperparameter settings for the maximum per-layer lookahead span $\forwardspan$ (i.e., 2, 5, and 10).
We report Word Error Rate (WER) (\%) and the statistical algorithmic latency (ms) of test utterances.
We also report on latency with median algorithmic latency and 90th percentile metrics, denoted as Alg.Latency@50 and Alg.Latency@90.
Included is the $\ell_1$-norm of the accessible future frames in attention maps to characterize the additional total amount of forward attention calculations (additional computation over causal).
The aggregation means are taken across all utterances in the test set.

Results for LibriSpeech are arranged in Table \ref{table:alg_main}.
While the algorithmic latency values for the non-adaptive models are static for a given $\forwardspan$, the WER and latency trade-off can be tuned by setting different penalty degrees on the regularizing term $\lambda$ for \acrshort{architecturename} models. 
To make clear comparisons of WER in Table \ref{table:alg_main}, we choose $\lambda$ values so that under each setting of $\forwardspan$ the Alg.Latency approximately matches that of the most efficient comparable baselines. 
As can be seen, training with our proposed \acrshort{architecturename} using our novel algorithmic latency loss as the regularizer, \acrshort{architecturename}\textit{-Alg.Lat.}, consistently yields lower WER over the approaches for equivalent latency operating points.
This holds for both Conformer and T-T encoders and all settings $\forwardspan$, seeing an average of 11\% (and upwards of 18\% for the lower latency scenarios) relative WER improvement over non-adaptive models.
Observe also, that compared with naive L1 regularization, for a given algorithmic latency budget, our \acrshort{architecturename}\textit{-Alg.Lat.} models allow for a greater total amount of future frame attention based on $\ell_1$-norm measures.
So while \acrshort{architecturename}\textit{-Alg.Lat.} uses overall higher compute, it is more strategic in how it expends it to deliver better WER for matching latency.

\begin{figure}[t!]
\centering
\includegraphics[width=0.95\columnwidth]{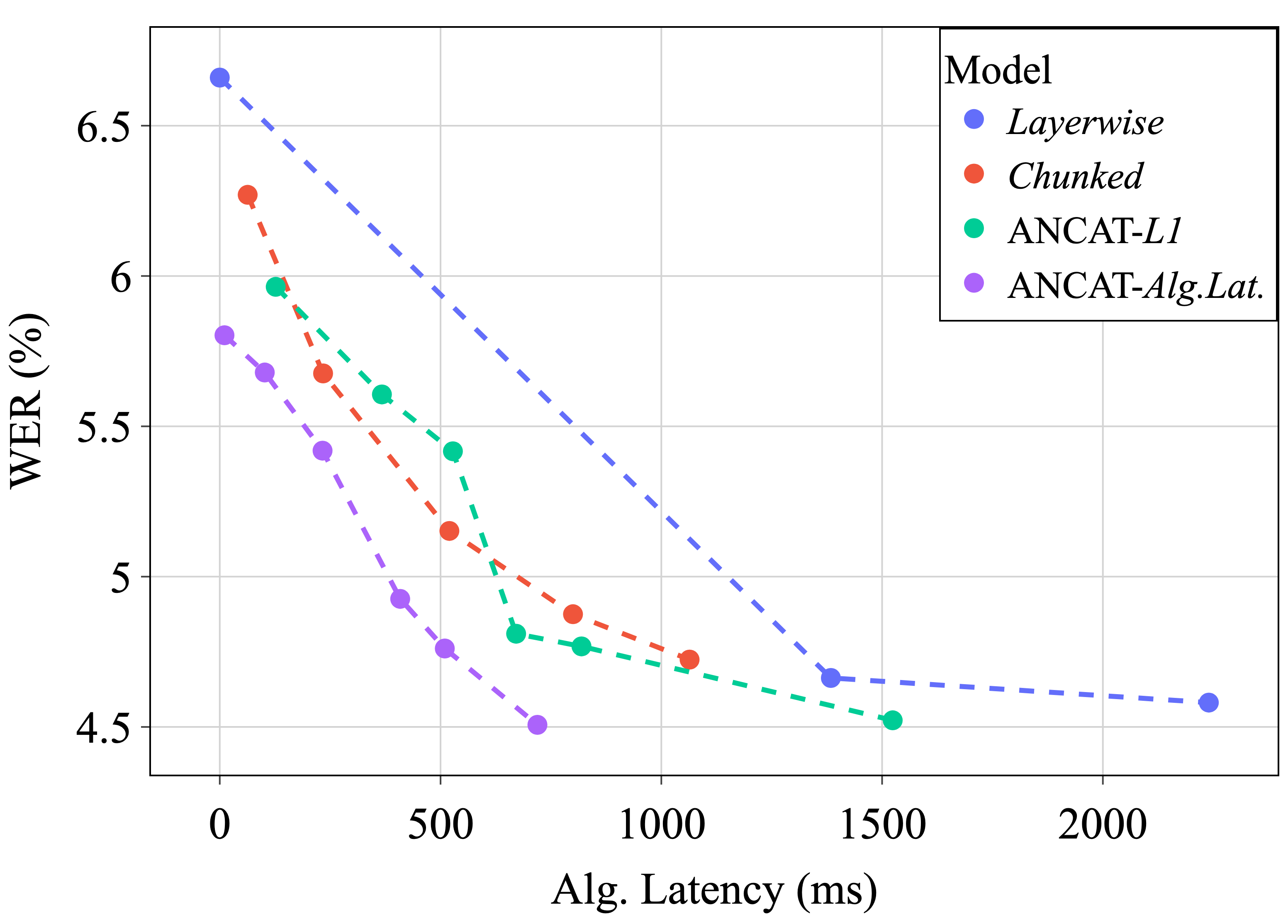}
\vspace{-3mm}
\caption{WER vs. algorithmic latency on LibriSpeech test ``clean'' data for four different types of Conformer models: \textit{Layerwise} ($\forwardspan$ at 0, 1, and 2 frames of right context per layer), \textit{Chunked} ($\forwardspan$ at 2, 5, 10, 15, and 20 frames), \acrshort{architecturename}\textit{-L1} ($\forwardspan=10$) varying $\lambda$, and \acrshort{architecturename}\textit{-Alg.Lat.} ($\forwardspan=10$) varying $\lambda$.
\acrshort{architecturename}\textit{-Alg.Lat.} provides a more optimal 
accuracy-latency trade-off over other models.}
\label{fig:algvswer}
\vspace{-0.3cm}
\end{figure}
To further highlight the improvements in the trade-off between WER and algorithmic latency, we record Conformer \acrshort{architecturename} model performance at multiple levels of training regularization in Figure~\ref{fig:algvswer}.
The plot shows that \acrshort{architecturename}\textit{-Alg.Lat.} consistently outperforms all other architectures at each operating point, and therefore fully defines the Pareto frontier of efficient solutions, providing the most optimal trade-offs of those models under consideration. 
Furthermore, the plot emphasizes that non-causal adaptivity alone \textit{does not} provide this improvement since \acrshort{architecturename}\textit{-L1} closely matches static \textit{Chunked} performance; rather, \acrshort{architecturename} adaptivity paired with the proper choice of our novel regularization method is critical to the success of the approach under latency considerations.

\begin{figure*}[t!]
\centering
\includegraphics[width=1.55\columnwidth]{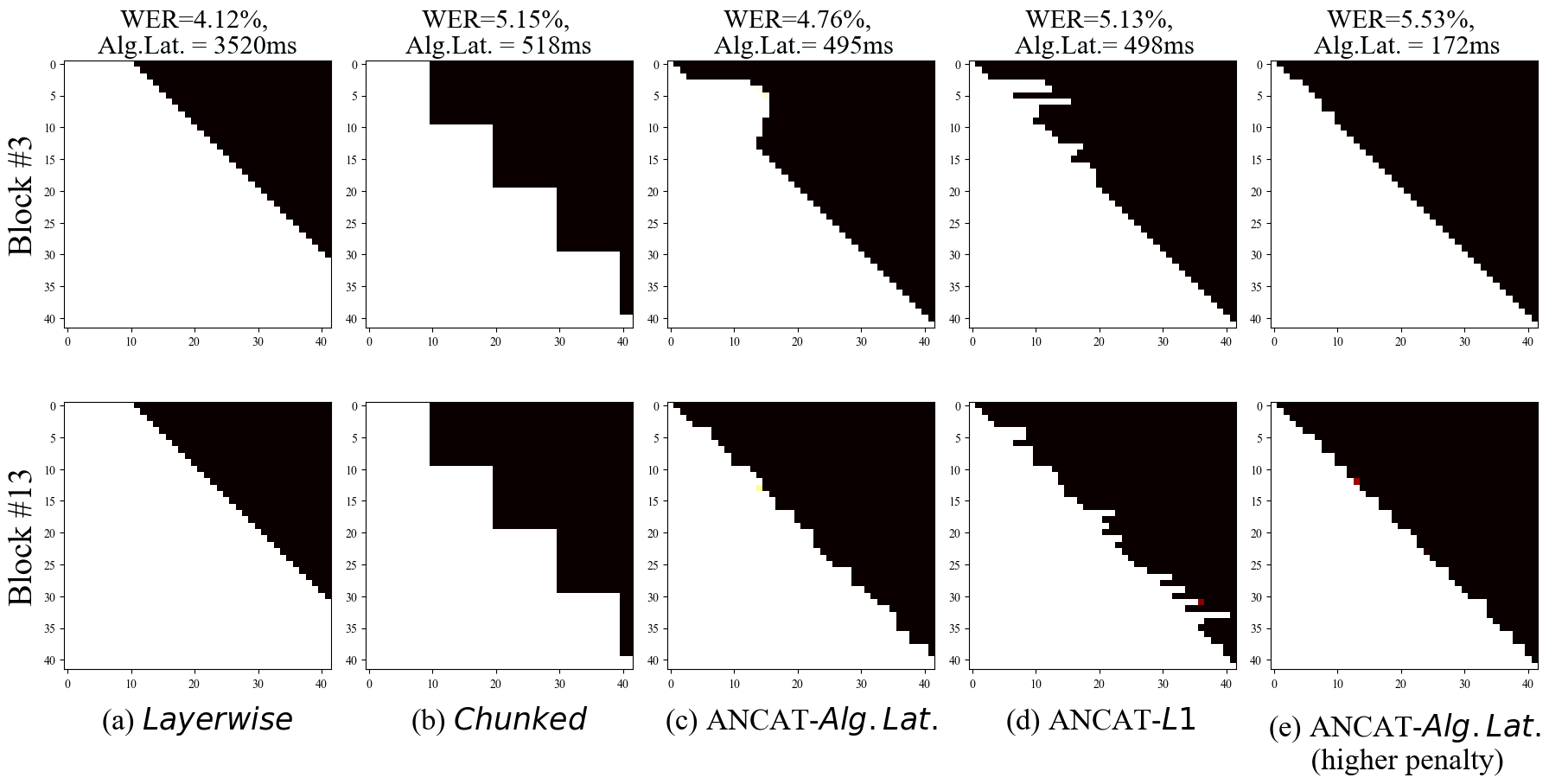}
\vspace{-4.5mm}
\caption{Attention mask visualization for the 3rd and 13th blocks of Conformer models.
All the attention masks are generated with the same test example utterance.
This utterance has a total number of 41 frames after downsampling in the front-end.
The max number of lookahead frames is $\forwardspan=10$.
The darker regions signal toggled-off attention.
We select (a), (b), and (c) models to have comparable algorithmic latency performance, and (d) shows a model that is optimized with a higher latency penalty.
Notice the deliberate, structured stepwise patterns \acrshort{architecturename}\textit{-Alg.Lat.} emits (c,e) compared to the jagged schedules produce by L1 regularization (d).}
\vspace{-2mm}
\label{fig:attention}
\end{figure*}
For further insight into the behavior of our proposed algorithmic latency loss and its impact on scheduling, we visualize the attention masks of an utterance from the test set in Figure~\ref{fig:attention} and compare against the static approaches.
The attention masks are generated with the final training epoch where $\tempparam=1e-4$.
The darker areas indicate the absence of attention.
We depict the maps from the 3rd block and the 13th block in the Conformer models.
One can observe that \acrshort{architecturename}\textit{-Alg.Lat.} learns to structure its attention in step-wise patterns which effectively operates like a chunking strategy of varying sizes and locations conditioned on the input.
In contrast, \acrshort{architecturename}\textit{-L1} attention patterns are jagged and emphasizes more localized decisions of the schedulers as opposed to the better coordinated decisions of \acrshort{architecturename}\textit{-Alg.Lat.} accounting for the global impact on latency.
We also show masks of an \acrshort{architecturename} model trained with higher latency penalization to demonstrate greater toggling-off of future attention connections, bordering being a fully causal model.

\subsection{Industrial Voice Assistant}
We repeat our algorithmic latency experiments on a de-identified large, industrial voice assistant dataset for again Conformer and T-T to verify robustness of our modeling approach both across architectures and data compositions.
We find similar results to that of LibriSpeech with \acrshort{architecturename}\textit{-Alg.Lat.} consistently outperforming \textit{Layerwise}, \textit{Chunked}, and \acrshort{architecturename}\textit{-L1} models at each operating point and improving upon WER by often greater than $5\%$ relative.
Appendix \ref{appendix:va_alg_lat} details this analogous experimental setup and its results.

\subsection{Compute-Induced UPL}
In addition to applying our algorithmic latency loss as the regularization method, we also conduct experiments to regularize with respect to compute-induced UPL, both for LibriSpeech and industrial data.
These results are presented in Appendix~\ref{appendix:upl} and mirror our findings with algorithmic latency, with \acrshort{architecturename} regularized with the backlog latency method presented in Section~\ref{subsec:computeuplloss} outperforming the baselines.
We find that for all compute throughput settings on which we experimented, for a given UPL budget, \acrshort{architecturename} delivers superior accuracy,
with typically \acrshort{architecturename}\textit{-UPL} improving WER by over 8\% relative on LibriSpeech and 7\% on industrial data over the nearest baselines.

\subsection{Additional Findings and Analysis}
We present supplemental visualizations and observations of the approach in further appendices. Appendix~\ref{appendix:el_ep_latency}
demonstrates that \acrshort{architecturename} promotes superior performance over baselines when comparing with other common latency metrics. Appendix~\ref{appendix:librispeech_other} shows that \acrshort{architecturename} also performs well (10\% WER relative improvement compared with baselines) in more challenging conditions using LibriSpeech ``other'' data and additive noise.
We further show how the difficulty of an utterance correlates to the degree of lookahead employed.
Appendix~\ref{appendix:attn_dynamics} highlights and expands upon additional attention plot examples while Appendix~\ref{appendix:training} depicts how the characteristics of \acrshort{architecturename} evolve over the course of training by temperature annealing.
The figures show that our ``soft'' attention connections and latency measures smoothly converge throughout training with the binary edge runtime latency calculations.

\vspace{-2mm}
\section{Conclusion}
\vspace{-1mm}
In this work, we introduce an adaptive non-causal architecture for streaming speech recognition.
The model learns to dynamically adjust the future context attention span for each individual frame of the audio stream, balancing both accuracy and latency considerations.
We accompanied our architecture construction with novel regularizing loss functions which tie the frame-wise lookahead decisions of the model with key latency measures important for speech applications.
The resulting mechanism provides a better Pareto frontier of trade-offs against baselines, in many cases with over 15\% relative WER improvements for matching latency.
Our experiments on public and large, production datasets and different architectures reinforce the robustness and applicability of our approach.
We hope that future work will build on these contributions to propose adaptive non-causal approaches for other applications, measures of latency, and models of computation.

\extrafootertext{The authors would like to thank Gautam Tiwari for his expert guidance and recommendations.}


\bibliography{example_paper}

\begin{thebibliography}{44}
\providecommand{\natexlab}[1]{#1}
\providecommand{\url}[1]{\texttt{#1}}
\expandafter\ifx\csname urlstyle\endcsname\relax
  \providecommand{\doi}[1]{doi: #1}\else
  \providecommand{\doi}{doi: \begingroup \urlstyle{rm}\Url}\fi

\bibitem[Audhkhasi et~al.(2021)Audhkhasi, Chen, Ramabhadran, and
  Moreno]{Audhkhasi2021}
Audhkhasi, K., Chen, T., Ramabhadran, B., and Moreno, P.~J.
\newblock {Mixture model attention: Flexible streaming and non-streaming
  automatic speech recognition}.
\newblock \emph{Proceedings of the Annual Conference of the International
  Speech Communication Association (INTERSPEECH)}, pp.\  4096--4100, 2021.

\bibitem[Bolukbasi et~al.(2017)Bolukbasi, Wang, Dekel, and
  Saligrama]{Bolukbasi2017}
Bolukbasi, T., Wang, J., Dekel, O., and Saligrama, V.
\newblock {Adaptive neural networks for efficient inference}.
\newblock \emph{International Conference on Machine Learning (ICML)}, pp.\
  812--821, 2017.

\bibitem[Chan et~al.(2016)Chan, Jaitly, Le, and Vinyals]{Chan2016}
Chan, W., Jaitly, N., Le, Q.~V., and Vinyals, O.
\newblock {Listen, attend and spell}.
\newblock \emph{IEEE International Conference on Acoustics, Speech and Signal
  Processing (ICASSP)}, pp.\  4960--4964, 2016.

\bibitem[Chang et~al.(2020)Chang, Subramanian, Guo, Watanabe, Fujita, and
  Omachi]{Chang2020}
Chang, X., Subramanian, A.~S., Guo, P., Watanabe, S., Fujita, Y., and Omachi,
  M.
\newblock {End-to-end ASR with adaptive span self-attention}.
\newblock \emph{Proceedings of the Annual Conference of the International
  Speech Communication Association (INTERSPEECH)}, pp.\  3595--3599, 2020.

\bibitem[Chen et~al.(2021)Chen, Wu, Wang, Liu, and Li]{Chen2021}
Chen, X., Wu, Y., Wang, Z., Liu, S., and Li, J.
\newblock {Developing real-time streaming transformer transducer for speech
  recognition on large-scale dataset}.
\newblock \emph{IEEE International Conference on Acoustics, Speech and Signal
  Processing (ICASSP)}, pp.\  5904--5908, 2021.

\bibitem[Chorowski et~al.(2015)Chorowski, Bahdanau, Serdyuk, Cho, and
  Bengio]{Chorowski2015}
Chorowski, J., Bahdanau, D., Serdyuk, D., Cho, K., and Bengio, Y.
\newblock {Attention-based models for speech recognition}.
\newblock \emph{Advances in Neural Information Processing Systems (NeurIPS)},
  pp.\  577--585, 2015.

\bibitem[Dehghani et~al.(2019)Dehghani, Gouws, Vinyals, Uszkoreit, and
  Kaiser]{Dehghani2019}
Dehghani, M., Gouws, S., Vinyals, O., Uszkoreit, J., and Kaiser, {\L}.
\newblock {Universal transformers}.
\newblock \emph{International Conference on Learning Representations (ICLR)},
  pp.\  1--23, 2019.

\bibitem[Dong et~al.(2019)Dong, Wang, and Xu]{Dong2019}
Dong, L., Wang, F., and Xu, B.
\newblock {Self-attention aligner: A latency-control end-to-end model for ASR
  using self-attention network and chunk-hopping}.
\newblock \emph{IEEE International Conference on Acoustics, Speech and Signal
  Processing (ICASSP)}, pp.\  5656--5660, 2019.

\bibitem[Elbayad et~al.(2020)Elbayad, Gu, Grave, and Auli]{Elbayad2020}
Elbayad, M., Gu, J., Grave, E., and Auli, M.
\newblock {Depth-adaptive transformer}.
\newblock \emph{International Conference on Learning Representations (ICLR)},
  2020.

\bibitem[Figurnov et~al.(2017)Figurnov, Collins, Zhu, Zhang, Huang, Vetrov, and
  Salakhutdinov]{Figurnov2017}
Figurnov, M., Collins, M.~D., Zhu, Y., Zhang, L., Huang, J., Vetrov, D., and
  Salakhutdinov, R.
\newblock {Spatially adaptive computation time for residual networks}.
\newblock \emph{IEEE Conference on Computer Vision and Pattern Recognition
  (CVPR)}, pp.\  1790--1799, 2017.

\bibitem[Graves(2012)]{Graves2012}
Graves, A.
\newblock {Sequence transduction with recurrent neural networks}.
\newblock \emph{International Conference on Machine Learning (ICML)}, 2012.

\bibitem[Graves(2016)]{Graves2016}
Graves, A.
\newblock {Adaptive computation time for recurrent neural networks}.
\newblock \emph{arXiv preprint arXiv:1603.08983}, 2016.

\bibitem[Gulati et~al.(2020)Gulati, Qin, Chiu, Parmar, Zhang, Yu, Han, Wang,
  Zhang, Wu, and Pang]{Gulati2020}
Gulati, A., Qin, J., Chiu, C.~C., Parmar, N., Zhang, Y., Yu, J., Han, W., Wang,
  S., Zhang, Z., Wu, Y., and Pang, R.
\newblock {Conformer: convolution-augmented transformer for speech
  recognition}.
\newblock \emph{Proceedings of the Annual Conference of the International
  Speech Communication Association (INTERSPEECH)}, pp.\  5036--5040, 2020.

\bibitem[Han et~al.(2020)Han, Zhang, Zhang, Yu, Chiu, Qin, Gulati, Pang, and
  Wu]{Han2020}
Han, W., Zhang, Z., Zhang, Y., Yu, J., Chiu, C.~C., Qin, J., Gulati, A., Pang,
  R., and Wu, Y.
\newblock {ContextNet: Improving convolutional neural networks for automatic
  speech recognition with global context}.
\newblock \emph{Proceedings of the Annual Conference of the International
  Speech Communication Association (INTERSPEECH)}, pp.\  3610--3614, 2020.

\bibitem[Han et~al.(2021)Han, Huang, Song, Yang, Wang, and Wang]{Han2021}
Han, Y., Huang, G., Song, S., Yang, L., Wang, H., and Wang, Y.
\newblock {Dynamic neural networks: A survey}.
\newblock \emph{IEEE Transactions on Pattern Analysis and Machine
  Intelligence}, 44\penalty0 (11):\penalty0 7436--7456, 2021.

\bibitem[He \& al.(2019)He and al.]{He2018}
He, Y. and al., E.
\newblock {Streaming end-to-end speech recognition for mobile devices}.
\newblock \emph{IEEE International Conference on Acoustics, Speech and Signal
  Processing (ICASSP)}, pp.\  6381--6385, 2019.

\bibitem[Inaguma et~al.(2020)Inaguma, Gaur, Lu, Li, and Gong]{Inaguma2020}
Inaguma, H., Gaur, Y., Lu, L., Li, J., and Gong, Y.
\newblock {Minimum latency training strategies for streaming
  sequence-to-sequence ASR}.
\newblock \emph{IEEE International Conference on Acoustics, Speech and Signal
  Processing (ICASSP)}, pp.\  6064--6068, 2020.

\bibitem[Jernite et~al.(2017)Jernite, Grave, Joulin, and Mikolov]{Jernite2017}
Jernite, Y., Grave, E., Joulin, A., and Mikolov, T.
\newblock {Variable computation in recurrent neural networks}.
\newblock \emph{International Conference on Learning Representations (ICLR)},
  pp.\  1--12, 2017.

\bibitem[Kingma \& {Lei Ba}(2015)Kingma and {Lei Ba}]{Kingma2015}
Kingma, D.~P. and {Lei Ba}, J.
\newblock {Adam: A method for stochastic optimization}.
\newblock \emph{International Conference on Learning Representations (ICLR)},
  2015.

\bibitem[Li et~al.(2021)Li, Gulati, Yu, Sainath, Chiu, Narayanan, Chang, Pang,
  He, Qin, Han, Liang, Zhang, Strohman, and Wu]{Li2021}
Li, B., Gulati, A., Yu, J., Sainath, T.~N., Chiu, C.~C., Narayanan, A., Chang,
  S.~Y., Pang, R., He, Y., Qin, J., Han, W., Liang, Q., Zhang, Y., Strohman,
  T., and Wu, Y.
\newblock {A better and faster end-to-end model for streaming ASR}.
\newblock \emph{IEEE International Conference on Acoustics, Speech and Signal
  Processing (ICASSP)}, pp.\  5634--5638, 2021.

\bibitem[Macoskey et~al.(2021{\natexlab{a}})Macoskey, Strimel, and
  Rastrow]{Macoskey2021a}
Macoskey, J., Strimel, G.~P., and Rastrow, A.
\newblock {Bifocal neural asr : exploiting keyword spotting for inference
  optimization}.
\newblock \emph{IEEE International Conference on Acoustics, Speech and Signal
  Processing (ICASSP)}, pp.\  5999--6003, 2021{\natexlab{a}}.

\bibitem[Macoskey et~al.(2021{\natexlab{b}})Macoskey, Strimel, Su, and
  Rastrow]{Macoskey2021}
Macoskey, J., Strimel, G.~P., Su, J., and Rastrow, A.
\newblock {Amortized neural networks for low-latency speech recognition}.
\newblock \emph{Proceedings of the Annual Conference of the International
  Speech Communication Association (INTERSPEECH)}, pp.\  1868--1872,
  2021{\natexlab{b}}.

\bibitem[Mohamed et~al.(2019)Mohamed, Okhonko, and Zettlemoyer]{Mohamed2019}
Mohamed, A., Okhonko, D., and Zettlemoyer, L.
\newblock {Transformers with convolutional context for ASR}.
\newblock \emph{arXiv preprint arXiv:1904.11660}, 2019.

\bibitem[Moritz et~al.(2020)Moritz, Hori, and Roux]{Moritz2020}
Moritz, N., Hori, T., and Roux, J.~L.
\newblock {Streaming automatic speech recognition with the transformer model}.
\newblock \emph{IEEE International Conference on Acoustics, Speech and Signal
  Processing (ICASSP)}, pp.\  6074--6078, 2020.

\bibitem[Moritz et~al.(2021)Moritz, Hori, and {Le Roux}]{Moritz2021}
Moritz, N., Hori, T., and {Le Roux}, J.
\newblock {Dual causal/non-causal self-attention for streaming end-to-end
  speech recognition}.
\newblock \emph{Proceedings of the Annual Conference of the International
  Speech Communication Association (INTERSPEECH)}, pp.\  4116--4120, 2021.

\bibitem[Narayanan et~al.(2021)Narayanan, Sainath, Pang, Yu, Chiu,
  Prabhavalkar, Variani, and Strohman]{Narayanan2021}
Narayanan, A., Sainath, T.~N., Pang, R., Yu, J., Chiu, C.~C., Prabhavalkar, R.,
  Variani, E., and Strohman, T.
\newblock {Cascaded encoders for unifying streaming and non-streaming ASR}.
\newblock \emph{IEEE International Conference on Acoustics, Speech and Signal
  Processing (ICASSP)}, pp.\  5629--5633, 2021.

\bibitem[Panayotov et~al.(2015)Panayotov, Chen, Povey, and
  Khudanpur]{Panayotov2015}
Panayotov, V., Chen, G., Povey, D., and Khudanpur, S.
\newblock {Librispeech: An ASR corpus based on public domain audio books}.
\newblock \emph{IEEE International Conference on Acoustics, Speech and Signal
  Processing(ICASSP)}, pp.\  5206--5210, 2015.

\bibitem[Park et~al.(2019)Park, Chan, Zhang, Chiu, Zoph, Cubuk, and
  Le]{Park2019}
Park, D.~S., Chan, W., Zhang, Y., Chiu, C.-c., Zoph, B., Cubuk, E.~D., and Le,
  Q.~V.
\newblock {Specaugment: A simple data augmentation method for automatic speech
  recognition}.
\newblock \emph{arXiv preprint arXiv:1904.08779}, 2019.

\bibitem[Peng et~al.(2023)Peng, Lee, and Watanabe]{Peng2023}
Peng, Y., Lee, J., and Watanabe, S.
\newblock {I3D: Transformer architectures with input-dependent dynamic depth
  for speech recognition}.
\newblock \emph{arXiv preprint arXiv:2303.07624}, 2023.

\bibitem[Radfar et~al.(2022)Radfar, Barnwal, Swaminathan, Chang, Strimel,
  Susanj, and Mouchtaris]{Radfar2022}
Radfar, M., Barnwal, R., Swaminathan, R.~V., Chang, F.~J., Strimel, G.~P.,
  Susanj, N., and Mouchtaris, A.
\newblock {ConvRNN-T: Convolutional Augmented Recurrent Neural Network
  Transducers for Streaming Speech Recognition}.
\newblock \emph{Proceedings of the Annual Conference of the International
  Speech Communication Association (INTERSPEECH)}, pp.\  4431--4435, 2022.

\bibitem[Sainath \& al.(2020)Sainath and al.]{Sainath2020}
Sainath, T.~N. and al., E.
\newblock {A streaming on-device end-to-end model surpassing server-side
  conventional model quality and latency}.
\newblock \emph{IEEE International Conference on Acoustics, Speech and Signal
  Processing (ICASSP)}, pp.\  6059--6063, 2020.

\bibitem[Shangguan et~al.(2021)Shangguan, Prabhavalkar, Su, Mahadeokar, Shi,
  Zhou, Wu, Le, Kalinli, Fuegen, and Seltzer]{Shangguan2021}
Shangguan, Y., Prabhavalkar, R., Su, H., Mahadeokar, J., Shi, Y., Zhou, J., Wu,
  C., Le, D., Kalinli, O., Fuegen, C., and Seltzer, M.~L.
\newblock {Dissecting user-perceived latency of on-device E2E speech
  recognition}.
\newblock \emph{arXiv preprint arXiv:2104.02207}, 2021.

\bibitem[Shi et~al.(2021)Shi, Wang, Wu, Yeh, Chan, Zhang, Le, and
  Seltzer]{Shi2021}
Shi, Y., Wang, Y., Wu, C., Yeh, C.-F., Chan, J., Zhang, F., Le, D., and
  Seltzer, M.
\newblock {Emformer: efficient nemory transformer based acoustic model for low
  latency streaming speech recognition}.
\newblock \emph{IEEE International Conference on Acoustics, Speech and Signal
  Processing (ICASSP)}, pp.\  6783--6787, 2021.

\bibitem[Song et~al.(2019)Song, Charlin, Xiao, Zhang, Wang, and Tang]{Song2019}
Song, W., Charlin, L., Xiao, Z., Zhang, M., Wang, Y., and Tang, J.
\newblock {Session-based social recommendation via dynamic graph attention
  networks}.
\newblock \emph{Proceedings of the Twelfth ACM International Conference on Web
  Search and Data Mining (WSDM)}, pp.\  555--563, 2019.

\bibitem[Sukhbaatar et~al.(2020)Sukhbaatar, Grave, Bojanowski, and
  Joulin]{Sukhbaatar2020}
Sukhbaatar, S., Grave, E., Bojanowski, P., and Joulin, A.
\newblock {Adaptive attention span in transformers}.
\newblock \emph{Annual Meeting of the Association for Computational Linguistics
  (ACL)}, pp.\  331--335, 2020.

\bibitem[Swietojanski et~al.(2022)Swietojanski, Braun, Can, da~Silva, Ghoshal,
  Hori, Hsiao, Mason, McDermott, Silovsky, Travadi, and
  Zhuang]{Swietojanski2022}
Swietojanski, P., Braun, S., Can, D., da~Silva, T.~F., Ghoshal, A., Hori, T.,
  Hsiao, R., Mason, H., McDermott, E., Silovsky, H., Travadi, R., and Zhuang,
  X.
\newblock {Variable attention masking for configurable transformer transducer
  speech recognition}.
\newblock \emph{arXiv preprint arXiv:2211.01438}, 2022.

\bibitem[Tripathi et~al.(2020)Tripathi, Kim, Zhang, Lu, and Sak]{Tripathi2020}
Tripathi, A., Kim, J., Zhang, Q., Lu, H., and Sak, H.
\newblock {Transformer transducer: One model unifying streaming and
  non-streaming speech recognition}.
\newblock \emph{arXiv preprint arXiv:2010.03192}, 2020.

\bibitem[Tsunoo et~al.(2019)Tsunoo, Kashiwagi, Kumakura, and
  Watanabe]{Tsunoo2019}
Tsunoo, E., Kashiwagi, Y., Kumakura, T., and Watanabe, S.
\newblock {Towards online end-to-end transformer automatic speech recognition}.
\newblock \emph{arXiv preprint arXiv:1910.11871}, 2019.

\bibitem[Wang et~al.(2020)Wang, Wu, Lu, Liu, Li, Ye, and Zhou]{Wang2020}
Wang, C., Wu, Y., Lu, L., Liu, S., Li, J., Ye, G., and Zhou, M.
\newblock {Low latency end-to-end streaming speech recognition with a scout
  network}.
\newblock \emph{Proceedings of the Annual Conference of the International
  Speech Communication Association (INTERSPEECH)}, pp.\  2112--2116, 2020.

\bibitem[Xie et~al.(2022)Xie, Macoskey, Radfar, Chang, King, Rastrow,
  Mouchtaris, and Strimel]{Xie2022}
Xie, Y., Macoskey, J., Radfar, M., Chang, F.~J., King, B., Rastrow, A.,
  Mouchtaris, A., and Strimel, G.~P.
\newblock {Compute Cost Amortized Transformer for Streaming ASR}.
\newblock \emph{Proceedings of the Annual Conference of the International
  Speech Communication Association (INTERSPEECH)}, pp.\  3043--3047, 2022.

\bibitem[Yeh et~al.(2019)Yeh, Mahadeokar, Kalgaonkar, Wang, Le, Jain, Schubert,
  Fuegen, and Seltzer]{Yeh2019}
Yeh, C.-F., Mahadeokar, J., Kalgaonkar, K., Wang, Y., Le, D., Jain, M.,
  Schubert, K., Fuegen, C., and Seltzer, M.~L.
\newblock {Transformer-transducer: End-to-end speech recognition with
  self-attention}.
\newblock \emph{arXiv preprint arXiv:1910.12977}, 2019.

\bibitem[Yu et~al.(2021{\natexlab{a}})Yu, Chiu, Li, Chang, Sainath, He,
  Narayanan, Han, Gulati, Wu, and Pang]{Yu2021}
Yu, J., Chiu, C.-C., Li, B., Chang, S.-y., Sainath, T.~N., He, Y., Narayanan,
  A., Han, W., Gulati, A., Wu, Y., and Pang, R.
\newblock {FastEmit: Low-latency streaming ASR with sequence-level emission
  regularization}.
\newblock \emph{IEEE International Conference on Acoustics, Speech and Signal
  Processing (ICASSP)}, pp.\  6004--6008, 2021{\natexlab{a}}.

\bibitem[Yu et~al.(2021{\natexlab{b}})Yu, Han, Gulati, Chiu, Li, Sainath, Wu,
  and Pang]{Yu2021a}
Yu, J., Han, W., Gulati, A., Chiu, C.-C., Li, B., Sainath, T.~N., Wu, Y., and
  Pang, R.
\newblock {Dual-mode ASR: Unify and improve streaming ASR with full-context
  modeling}.
\newblock \emph{International Conference on Learning Representations (ICLR)},
  2021{\natexlab{b}}.

\bibitem[Zhang et~al.(2020)Zhang, Lu, Sak, Tripathi, Mcdermott, Koo, and
  Kumar]{Zhang2020}
Zhang, Q., Lu, H., Sak, H., Tripathi, A., Mcdermott, E., Koo, S., and Kumar, S.
\newblock {Transformer transducer: a streamable speech recognition model with
  transformer encoders and RNN-T loss}.
\newblock \emph{IEEE International Conference on Acoustics, Speech and Signal
  Processing (ICASSP)}, pp.\  7829--7833, 2020.

\end{thebibliography}
\bibliographystyle{icml2023}

\newpage
\appendix
\onecolumn

\section{Emission Latency and Endpointer Latency}
\label{appendix:el_ep_latency}

To conduct a more comprehensive investigation into the improved latency performance resulting from the \acrshort{architecturename} architecture, we carry out measurements on additional emission/transcript (EM) latency and endpointer (EP) latency in this section. EM.Latency refers to the mean difference between the frame in which a token is produced by decoding (accounting for future lookahead) and when the token is spoken given by alignment data. EP.Latency refers specifically to the difference between when an end-of-speech token is emitted and the final word is spoken (again, accounting for future lookahead). In Table \ref{table:el_ep_latency}, the numbers show comparable (or better) EM.Latency and EP.Latency to the best baseline, while the WER of our proposed \acrshort{architecturename}-\textit{Alg.Lat.}\strut is superior.

\begin{table*}[ht!]
\caption{Results on LibriSpeech test ``clean'' for Conformer.
The models in this table are identical with the models in Table 1 in main body of the paper. We report the average emission/transcript (EM) latency and endpointer latency (EP) in this table. 
}

\centering
\scalebox{0.95}{
\begin{tabular}{c|p{0.5cm}p{0.5cm}p{0.5cm}p{0.5cm}p{0.5cm}p{0.5cm}p{0.5cm}p{0.5cm}p{0.5cm}p{0.5cm}p{0.5cm}p{0.5cm}p{0.5cm}}
\specialrule{.2em}{.1em}{.1em} 
\rule{0pt}{15pt}  & \multicolumn{13}{c}{Conformer}                   \\ 

\hline

\rule{0pt}{15pt} $\forwardspan$ ($\#$ lookahead) &\multicolumn{1}{c}{0}&\multicolumn{4}{c}{2}&     \multicolumn{4}{c}{5}&\multicolumn{4}{c}{10}       \\ 
\cmidrule(r{2pt}){1-1}\cmidrule(l{2pt}){2-2}\cmidrule(l){3-6}\cmidrule(l){7-10}\cmidrule(lr){11-14}\\

Model         &  
\multicolumn{1}{c}{\rotatebox[origin=l]{90}{\tikz\draw[color=red!80, fill=red!40] (0,0) circle (.5ex); \textit{Causal}\strut} }&  
\multicolumn{1}{c} {\rotatebox[origin=l]{90}{\tikz\draw[magenta!80,fill=magenta!20] (0,-1) circle (.5ex); \textit{Layerwise}\strut}}   & 
\multicolumn{1}{c} {\rotatebox[origin=l]{90}{\tikz\draw[magenta!80,fill=magenta!20] (0,0) circle (.5ex); \textit{Chunked}\strut}} & 
\multicolumn{1}{c} {\rotatebox[origin=l]{90}{\tikz\draw[magenta!80,fill=magenta!20] (0,0) circle (.5ex); \acrshort{architecturename}-\textit{L1}\strut}} &  
\multicolumn{1}{c} {\rotatebox[origin=l]{90}{\tikz\draw[magenta!80,fill=magenta!20] (0,0) circle (.5ex); \acrshort{architecturename}-\textit{Alg.Lat.}\strut}} &  
\multicolumn{1}{c} {\rotatebox[origin=l]{90}{\tikz\draw[color=orange!80, fill=orange!20] (0,0) circle (.5ex); \textit{Layerwise}\strut}} & 
 \multicolumn{1}{c} {\rotatebox[origin=l]{90}{\tikz\draw[color=orange!80, fill=orange!20] (0,0) circle (.5ex); \textit{Chunked}\strut}} & 
 \multicolumn{1}{c} {\rotatebox[origin=l]{90}{\tikz\draw[color=orange!80, fill=orange!20] (0,0) circle (.5ex); \acrshort{architecturename}-\textit{L1}\strut}} & 
 \multicolumn{1}{c} {\rotatebox[origin=l]{90}{\tikz\draw[color=orange!80, fill=orange!20] (0,0) circle (.5ex); \acrshort{architecturename}-\textit{Alg.Lat.}\strut}}& 
 \multicolumn{1}{c}{\rotatebox[origin=l]{90}{\tikz\draw[color=blue!80, fill=blue!20] (0,0) circle (.5ex);  \textit{Layerwise}\strut} }& 
 \multicolumn{1}{c}{\rotatebox[origin=l]{90}{\tikz\draw[color=blue!80, fill=blue!20] (0,0) circle (.5ex); \textit{Chunked}\strut} }& 
 \multicolumn{1}{c}{\rotatebox[origin=l]{90}{\tikz\draw[color=blue!80, fill=blue!20] (0,0) circle (.5ex); \acrshort{architecturename}-\textit{L1}\strut} }& 
\multicolumn{1}{c} {\rotatebox[origin=l]{90}{\tikz\draw[color=blue!80, fill=blue!20] (0,0) circle (.5ex); \acrshort{architecturename}-\textit{Alg.Lat.}\strut} }\\ 

\cmidrule(r{2pt}){1-1}\cmidrule(l{2pt}){2-2}\cmidrule(l){3-6}\cmidrule(l){7-10}\cmidrule(lr){11-14}\\
WER (\%)          & 6.66  & 4.58  & 5.68& 6.27   & \textbf{5.31} &  4.23 & 5.68 & 5.43&\textbf{5.13} & 4.12 & 5.15 &5.13 & \textbf{4.76}   \\
EM.Latency (ms)  & 235 & 2130 & 305  & 296 &\textbf{290} & 3125 & 395  & 315& \textbf{282} & 3418 & 622&560  & \textbf{551}  \\
EP.Latency (ms) & 74 & 1992 & 141 & 116 &\textbf{116}  &  2992 & 252  & 172&\textbf{139} &  3286 &490 & 402& \textbf{408} \\\hline
\end{tabular}}
\label{table:el_ep_latency}
\end{table*}

\section{LibriSpeech ``Other'' and Noise Impact}
\label{appendix:librispeech_other}
\begin{figure}[H]
\centering
\includegraphics[width=0.47\columnwidth]{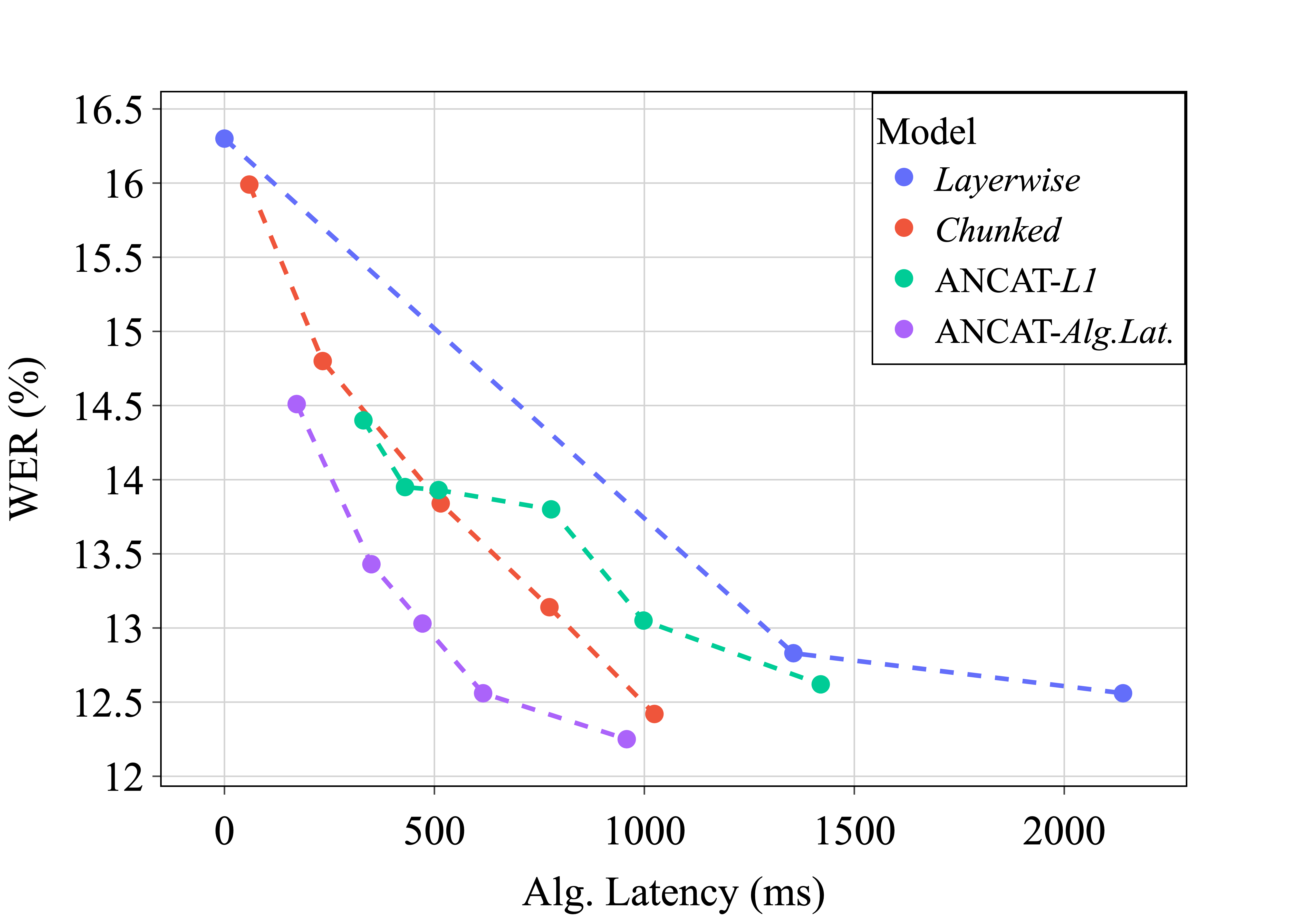}
\vspace{-3mm}
\caption{WER vs. algorithmic latency on LibriSpeech test ``other'' data for four different types of Conformer models: \textit{Layerwise} ($\forwardspan$ at 0, 1, and 2 frames of right context per layer), \textit{Chunked} ($\forwardspan$ at 2, 5, 10, 15, and 20 frames), \acrshort{architecturename}\textit{-L1} ($\forwardspan=10$) varying $\lambda$, and \acrshort{architecturename}\textit{-Alg.Lat.} ($\forwardspan=10$) varying $\lambda$.
\acrshort{architecturename}\textit{-Alg.Lat.} provides a more optimal 
accuracy-latency trade-off over other models.
}
\label{fig:libri_alg_v_wer}
\end{figure}

\begin{figure}[H]
\centering
\subfigure{
    \centering
    \includegraphics[width=0.47\columnwidth]{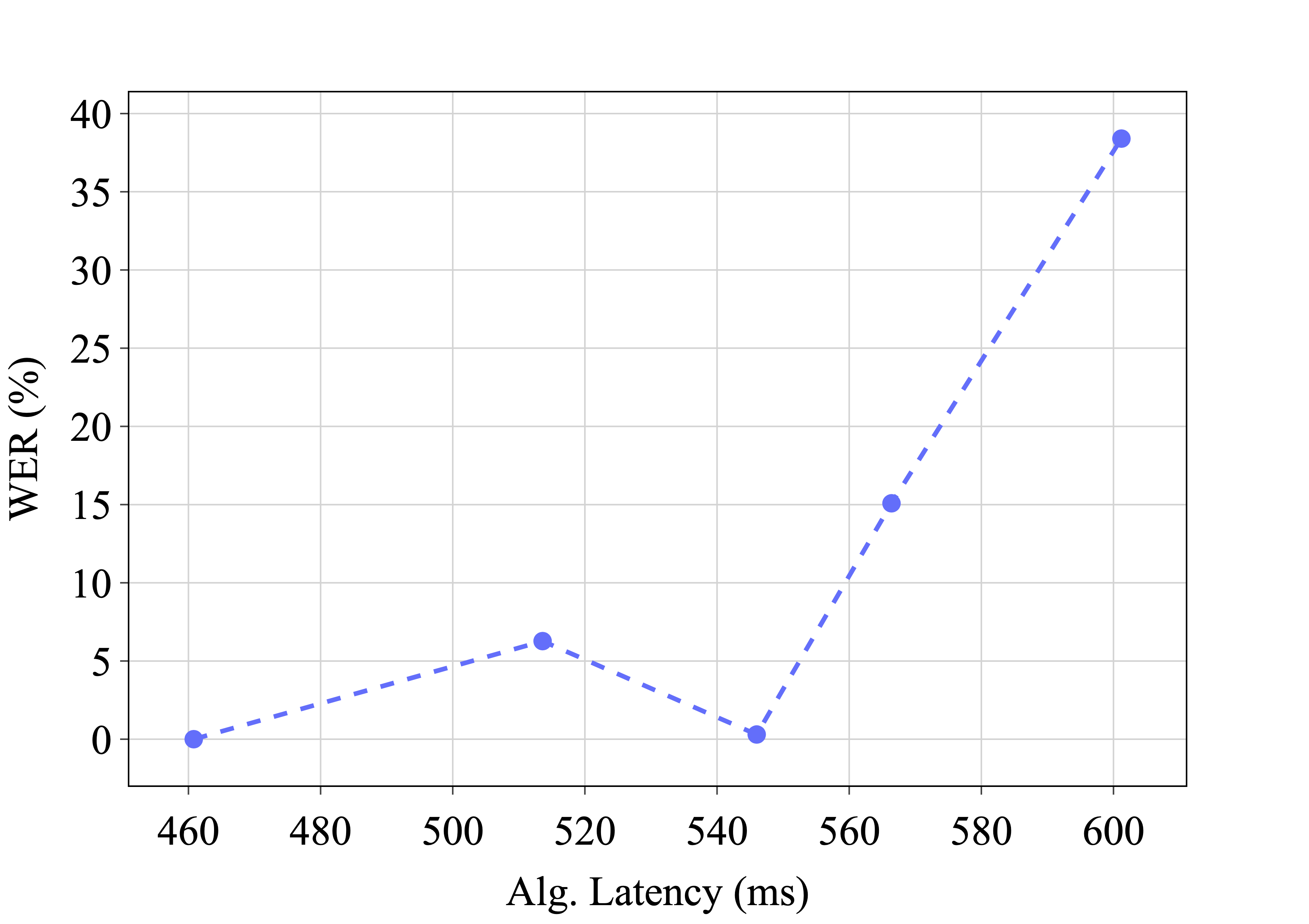}}%
\subfigure{
    \centering
    \includegraphics[width=0.47\columnwidth]{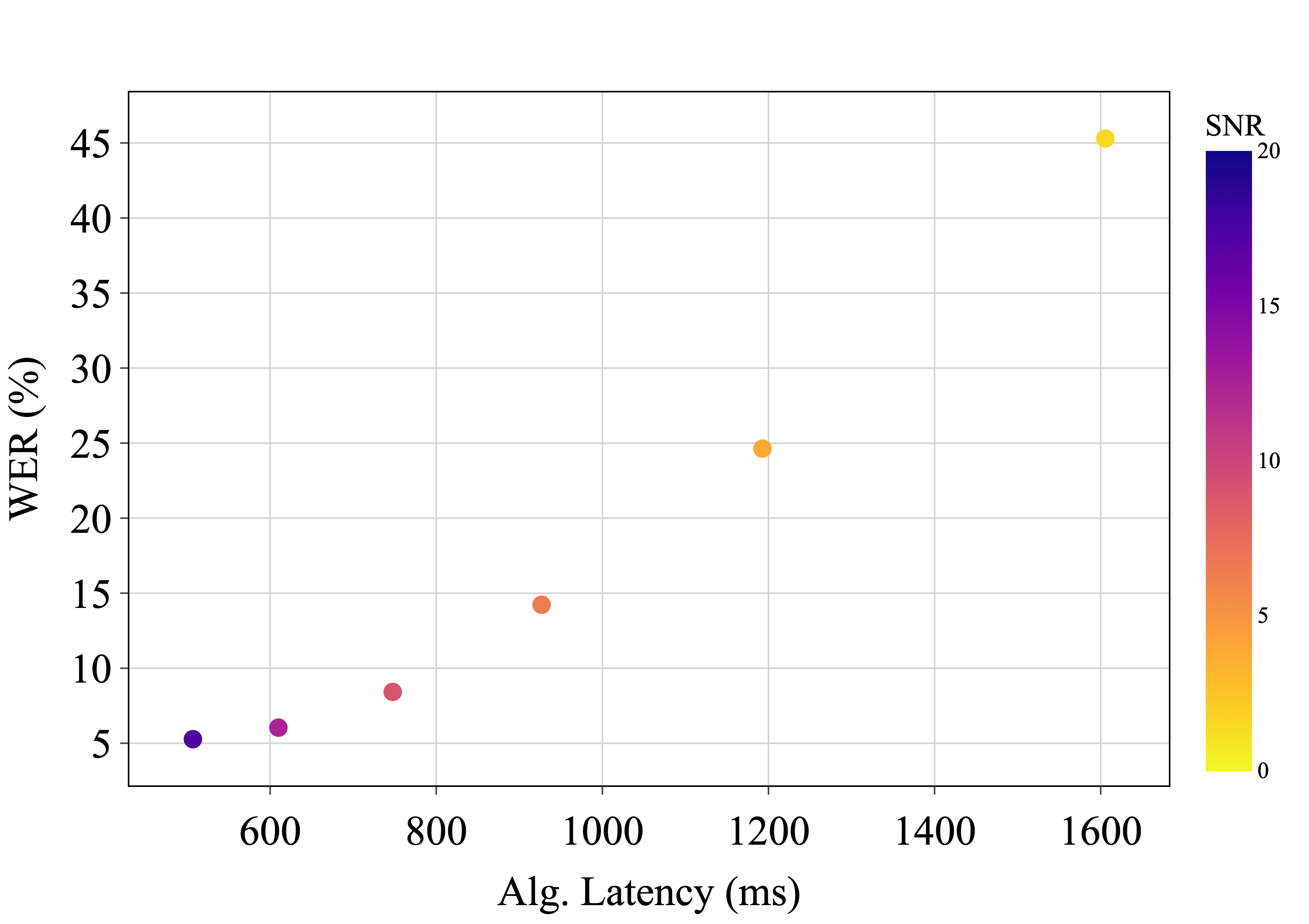}
}%
\vspace{-3mm}
\caption{WER vs. algorithmic latency for single \acrshort{architecturename}\textit{-Alg.Lat.} model. The left sub-figure is LibriSpeech test ``clean'' and ``other'' utterances are grouped into five percentiles based on individual WER with the aggregate WER within each group then plotted against its mean latency. The right sub-figure shows this same model performing on the test ``clean'' dataset with different levels of white noise added over the audio signals. There is a clear trend between the difficulty of the utterance and the amount of lookahead employed resulting in higher latency for more challenging examples.
}
\label{fig:libri_noise}
\end{figure}

The LibriSpeech corpus also provides the more difficult test ``other'' dataset comprised of 5.1 hours of speech signals chosen from more a challenging speaker cohort.
Results of our \acrshort{architecturename} and baseline models are presented in Figure~\ref{fig:libri_alg_v_wer}.
Besides the naturally higher error rates for all models, we observe a very similar plot to that of test ``clean'' with  \acrshort{architecturename}\textit{-Alg.Lat.} providing the most optimal accuracy-latency trade-offs. 
Again, \acrshort{architecturename}\textit{-Alg.Lat.} defines the Pareto frontier with respect to algorithmic latency and WER for those models considered while yielding upwards of 10\% relative WER improvements for several operating points. 
\vspace{-3.5mm}
\vspace{-3mm}
One observes that while WER increases for test ``other'' data, the algorithmic latency also increases noticeably as well.
This leads us to examine for a specified \acrshort{architecturename} model, the relationship between WER and algorithmic latency.
Do naturally more difficult utterances (producing higher WER) correlate with inflated algorithmic latency driven by \acrshort{architecturename} schedulers?
Figure~\ref{fig:libri_noise} (left) answers this in the affirmative.
The visualization is constructed by grouping all individual test ``clean'' and ``other'' utterance WERs into five percentiles.
The WER of each of these groups is then compared with its mean latency to observe a positive association between the two quantities.
This finding suggests that for more challenging utterances, \acrshort{architecturename} indeed will resort to using more lookahead attention.
We also show a simple experiment in Figure~\ref{fig:libri_noise} (right) which adds different degrees of noise to LibriSpeech test ``clean'' data and illustrates a smooth trend between latency and WER/noise. 
This finding supports the notion that under more challenging conditions, \acrshort{architecturename}\textit{-Alg.Lat.} models will opt to use more lookahead attention to compensate for the added ambiguity.
\vspace{-3.5mm}

\section{Voice Assistant Data}
\label{appendix:va_alg_lat}
In this section we discuss the experimental setup for Voice Assistant data and algorithmic latency results mirroring those of LibriSpeech. Further appendices will present additional findings and figures inclusive of both LibriSpeech and Voice Assistant experiments.
\subsection{Experimental Setup}
Our in-house datasets consist of de-identified, far-field, voice assistant utterances.
The training dataset consists of approximately 150k hours of transcribed audio.
We used model architectures identical to those we built for LibriSpeech, with just a couple of configuration differences better fitting for the voice assistant data.
Specifically, we used a slightly larger word piece tokenization model of size 4k and the two layer convolutional front-end had stride sizes of 2 and 1 (resulting in $60$ ms \acrshort{architecturename} frames) as opposed to 2 and 2 stride sizes for LibriSpeech.
All our accuracy numbers for internal data we report as Word Error Rate Relative (WERR) with respect to the baseline fully causal model (below 7.5\% WER absolute).

\subsection{Algorithmic Latency}
Plots for Conformer results pertaining to algorithmic latency are presented in the following figures.
We couple figures with a final table in Appendix~\ref{appndx:tables} for reference.
Figure~\ref{fig:alexa_alg_v_wer} visualizes how WER and algorithmic latency vary between the different types of models and different
regularization factors.
As done for Figure~\ref{fig:algvswer}, we adjust the lookahead span for the static models \textit{Layerwise} and \textit{Chunked} while fixing $\forwardspan$ at 10 frames per layer for the \acrshort{architecturename} models and adjusting their regularization penalty.
These results match closely with those of LibriSpeech with \acrshort{architecturename}\textit{-Alg.Lat.} providing the most optimal accuracy-latency operating points.
In many cases, \acrshort{architecturename}\textit{-Alg.Lat.} outperforms its nearest competitor by 5\% relative in WER for matching latency budgets.
Again, one finds that \acrshort{architecturename}\textit{-L1} performs comparable with \textit{Chunked} and demonstrates that the novel latency regularization algorithm is vital to the adaptive approach's success over that of a more naive regularization scheme.
\vspace{5mm}
\begin{figure}[H]
\centering
\includegraphics[width=0.65\columnwidth]{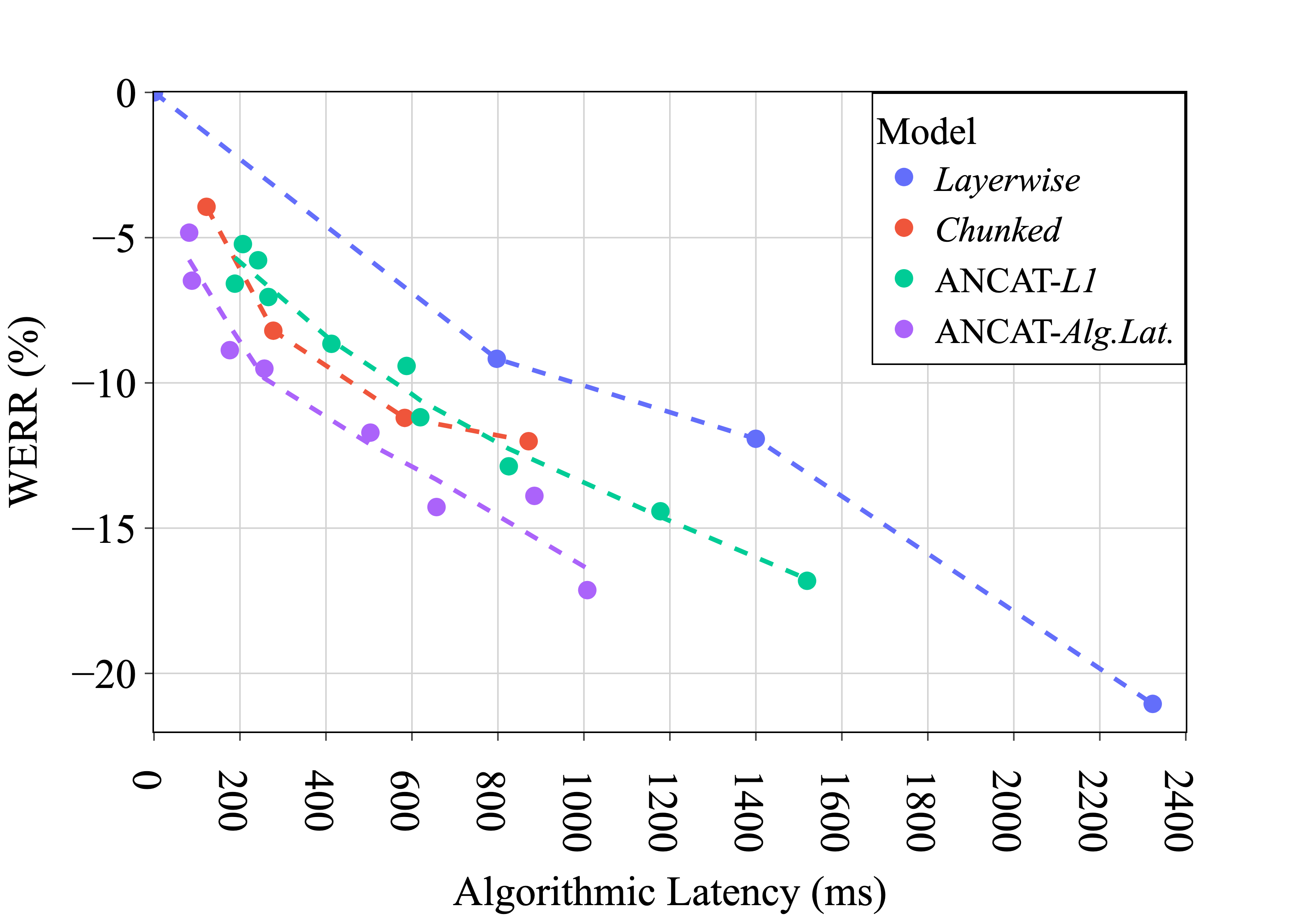}
\caption{WERR vs. algorithmic latency on Voice Assistant data for four different types of Conformer models: 1) \textit{Layerwise} ($\forwardspan$ at 0, 1, 2,
and 10 frames of right context per layer), \textit{Chunked} ($\forwardspan$ at 5, 10, 20, and 30 frames), \acrshort{architecturename}\textit{-L1} ($\forwardspan=10$) varying regularization term $\lambda$, and \acrshort{architecturename}\textit{-Alg.Lat.} ($\forwardspan=10$) varying $\lambda$.
}
\label{fig:alexa_alg_v_wer}
\end{figure}
\vspace{2.5mm}

\begin{figure}[H]
\centering
\includegraphics[width=0.9\columnwidth]{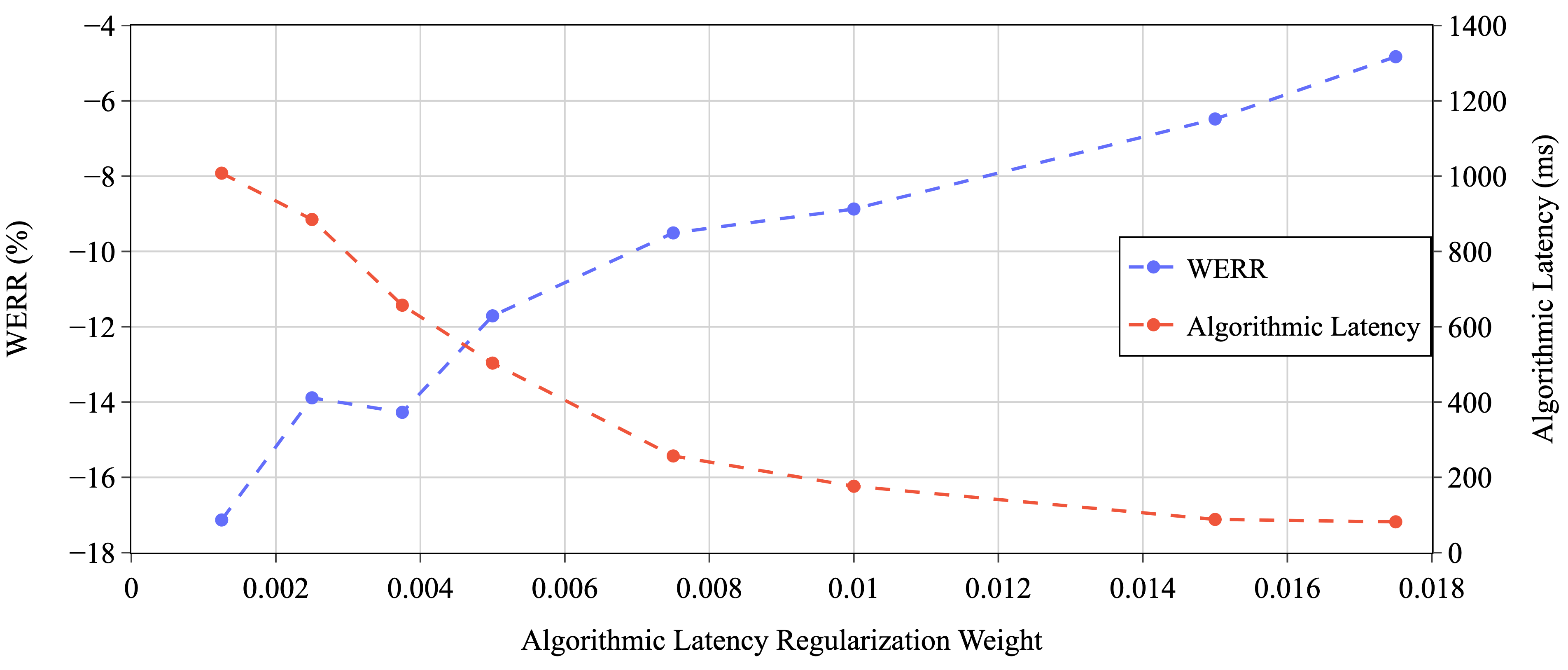}
\caption{WERR and algorithmic latency vs. \acrshort{architecturename} regularization weight $\lambda$ for a Conformer \acrshort{architecturename}\textit{-Alg.Lat.} model on Voice Assistant data.}
\label{fig:alexa_werr_aa_regweight}
\end{figure}

To complement Figure~\ref{fig:alexa_alg_v_wer} above and emphasise the tuneability of \acrshort{architecturename}, Figure~\ref{fig:alexa_werr_aa_regweight} shows how WER and latency behave as a function of the regularization constant $\lambda$ for Conformer \acrshort{architecturename}\textit{-Alg.Lat.} model on Voice Assistant data.
From this figure, one can see that the WERR and latency curves are relatively smooth over the weights, so it is easy to select a desired accuracy-latency operating point.
Finally, note that \acrshort{architecturename} models with zero to high regularization are indeed close to the operating points expected: near zero latency for high regularization and significant accuracy improvements similar to maximal right context models for minimal regularization.

\section{Compute-Induced UPL Results}
\label{appendix:upl}
Section~\ref{subsec:computeuplloss} provides a novel regularization method to account for both the processing power of executing hardware along with the schedulers' decisions in order to reduce the approximate UPL an application might expect with an \acrshort{architecturename} model.
Using this notion of latency, we experiment with LibriSpeech and Voice Assistant data to build \acrshort{architecturename}\textit{-UPL} models with UPL regularization.
For our 14-block model architecture and frame rate $\rho$, the experiments consider compute throughput values $\mu$ of 1.5, 2, and 3 ($\times 14\rho$).
The values represent the number of nodes in the compute DAG which can be processed per second and intuitively are equivalent to the speed of 1.5, 2 and 3 forward passes through the model per frame (i.e. real-time multipliers).

\begin{figure}[H]
\centering
\subfigure{
    \centering
    \includegraphics[width=0.5\columnwidth]{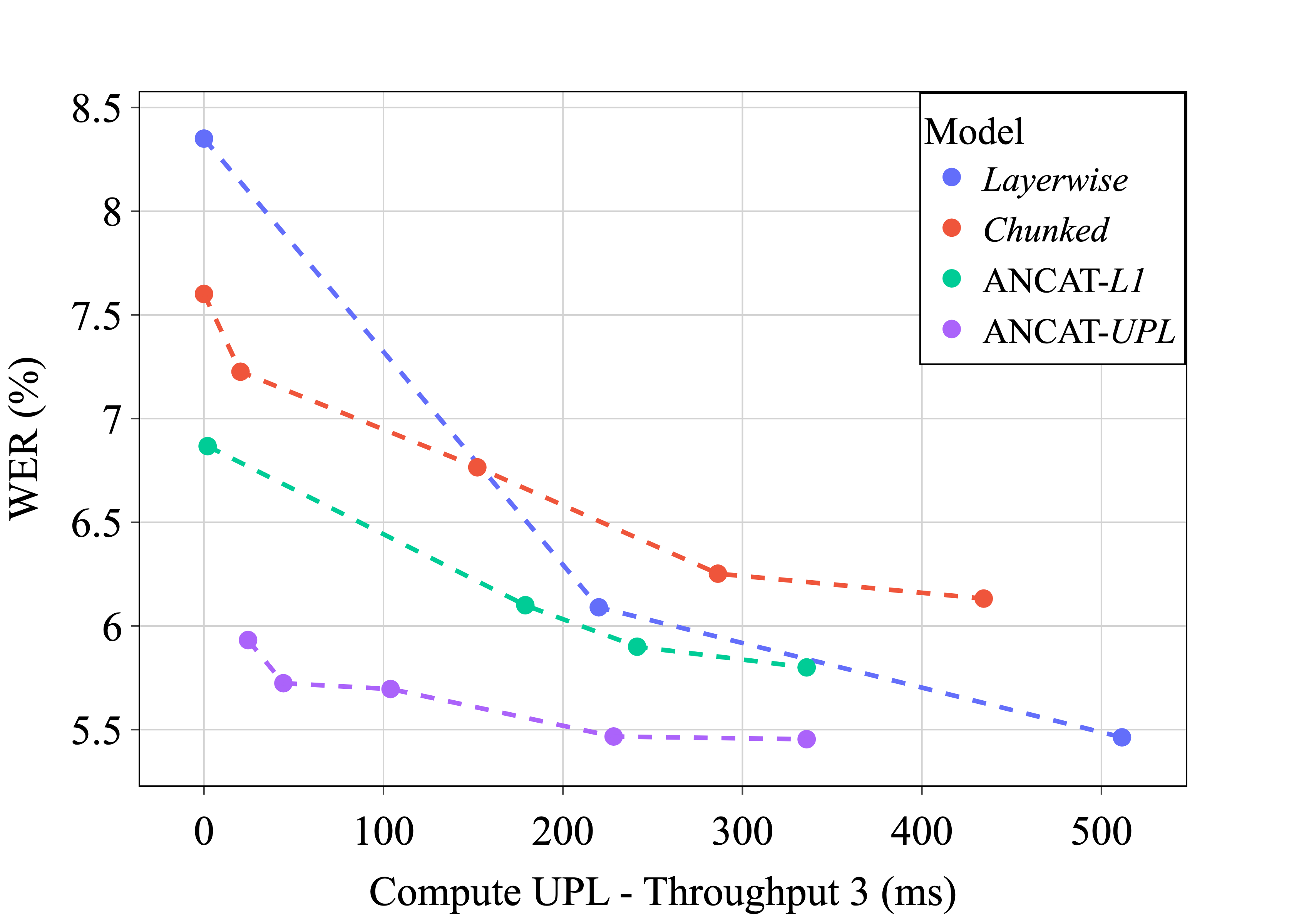}}%
\subfigure{
    \centering
    \includegraphics[width=0.5\columnwidth]{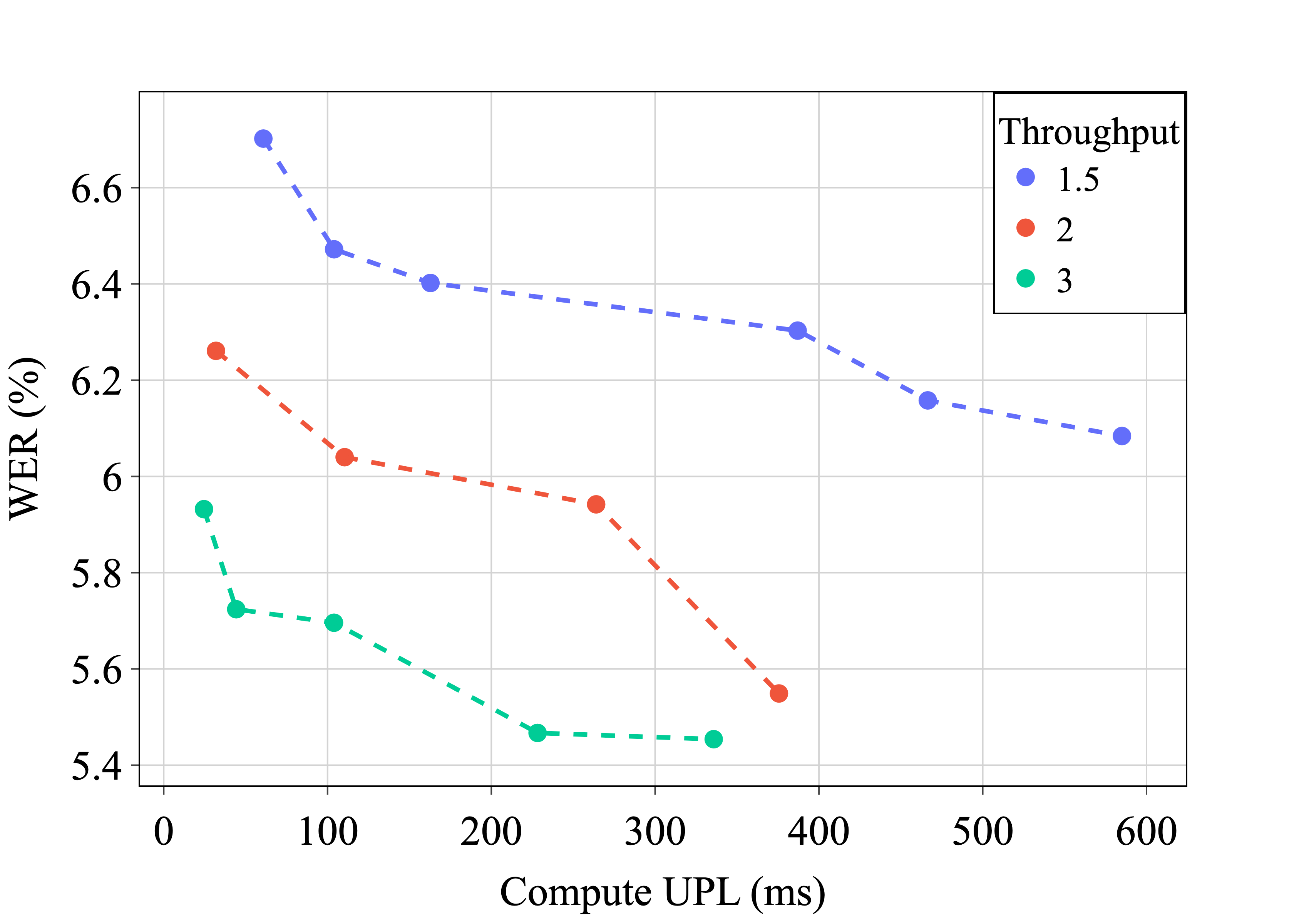}
}%
\caption{
LibriSpeech test ``clean'': Compute UPL vs. WER. (Left) For a fixed throughput, UPL from the compute backlog comparing four different types of Conformer models: 1) \textit{Layerwise} attention models ($\forwardspan$ at 0, 1, and 2 frames of right context per layer), 2) \textit{Chunked}  ($\forwardspan$ at 2, 5, 10, 15 and 20 frames of right context per layer), \acrshort{architecturename}\textit{-L1}, and 4) \acrshort{architecturename}\textit{-UPL} with \acrshort{architecturename} models having $\forwardspan=10$ and varying regularization term $\lambda$. With respect to perceived latency, \acrshort{architecturename}\textit{-UPL} provides the superior accuracy-latency trade-off. (Right) \acrshort{architecturename}\textit{-UPL} models trained with a fixed regularization factor but with different throughputs.
Faster compute power gives the model more lookahead flexibility, resulting in better accuracy-latency operating points for \acrshort{architecturename}\textit{-UPL} models.
From our empirical observations, the UPL loss behaves sensitively time masking patterns and is prone to overfitting on small datasets. Thereby, for UPL results we only apply frequency masking of the SpecAugmentation which leads a slightly higher WER over our other LibriSpeech experiments.}
\label{fig:libri_upl}
\end{figure}

\begin{figure}[H]
\centering
\subfigure{
    \centering
    \includegraphics[width=0.5\columnwidth]{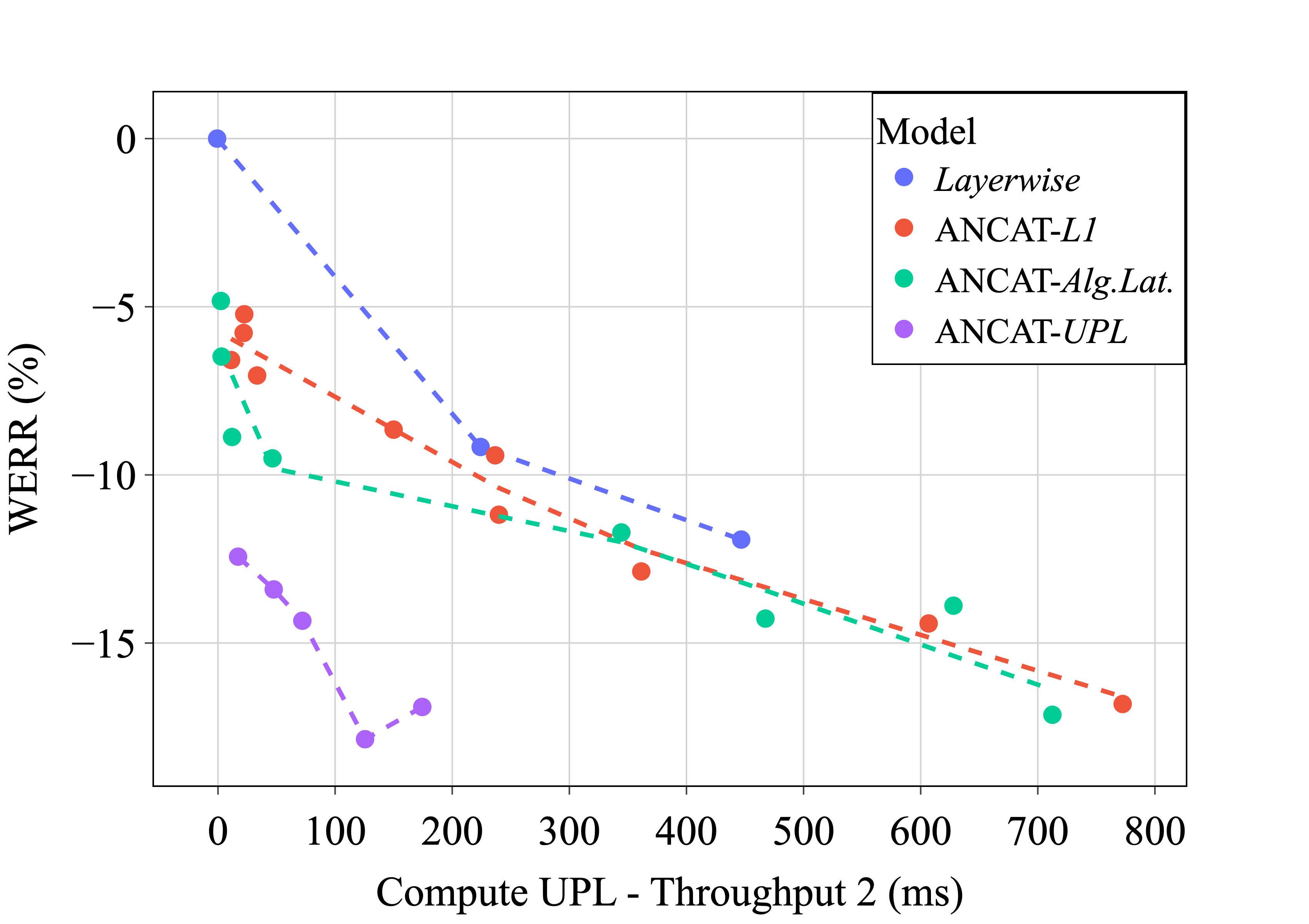}}%
\subfigure{
    \centering
    \includegraphics[width=0.5\columnwidth]{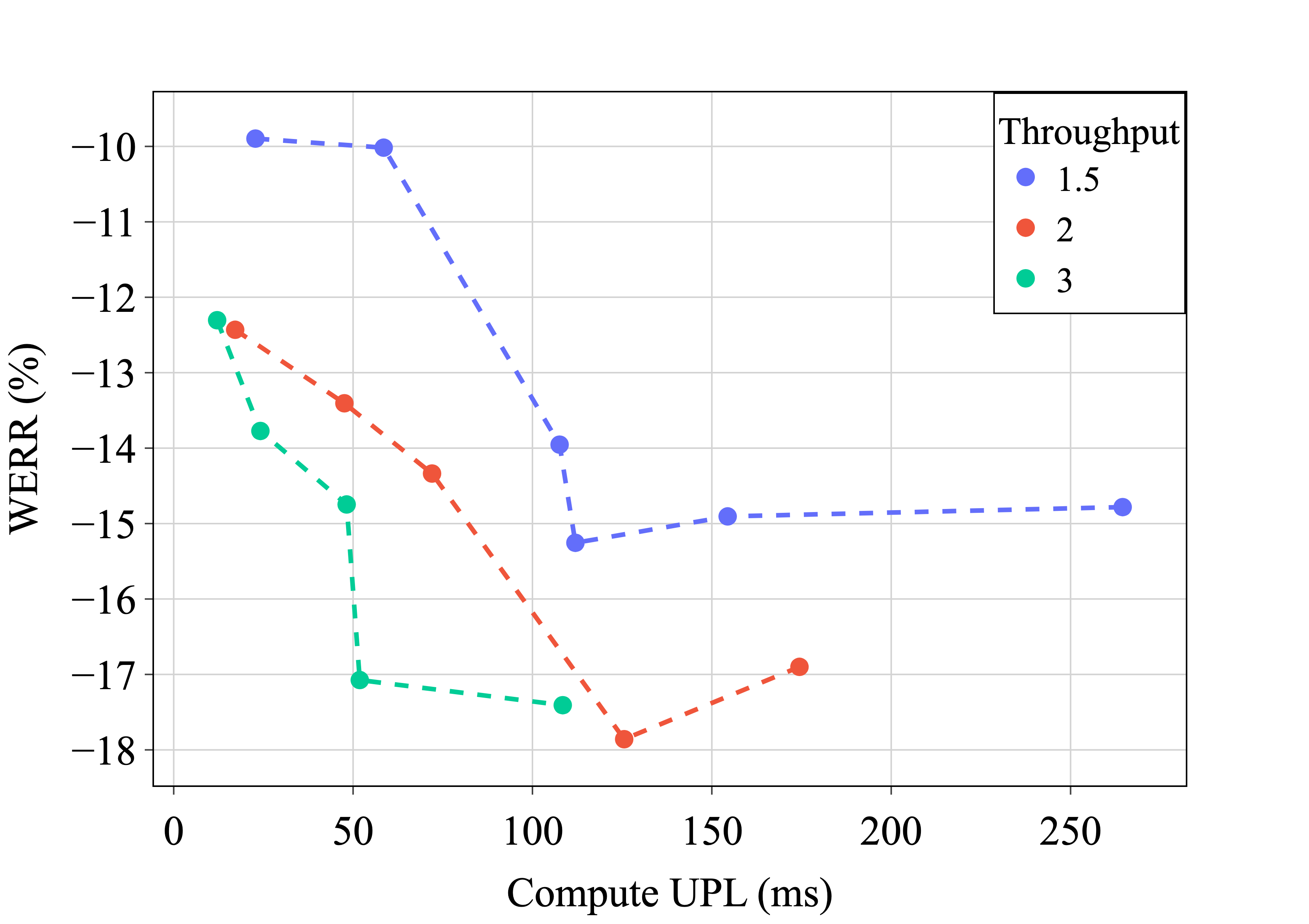}
}%
\caption{Voice Assistant: Compute UPL vs. WERR. (Left) For a fixed throughput, UPL from the compute backlog comparing four different types of Conformer models: 1) \textit{Layerwise} attention models ($\forwardspan$ at 0, 1, and 2 frames of right context per layer), 2) \acrshort{architecturename}\textit{-L1}, 3) \acrshort{architecturename}\textit{-Alg.Lat.}, and 4) \acrshort{architecturename}\textit{-UPL} with \acrshort{architecturename} models having $\forwardspan=10$ and varying regularization term $\lambda$. With respect to perceived latency, \acrshort{architecturename}\textit{-UPL} provides the superior accuracy-latency trade-off. (Right) \acrshort{architecturename}\textit{-UPL} models trained with a fixed regularization factor but with different throughputs.
Faster compute power gives the model more lookahead flexibility, resulting in better accuracy-latency operating points for \acrshort{architecturename}\textit{-UPL} models.
}\label{fig:va_upl}
\end{figure}

In Figures~\ref{fig:libri_upl} and \ref{fig:va_upl}, we see how WER and compute UPL vary between models with different regularization weights and compute throughput.
We see that for a fixed throughput value, as the regularization is decreased, the UPL smoothly increases, analogous to our algorithmic latency results.
Further we find that for both LibriSpeech and Voice Assistant, \acrshort{architecturename}\textit{-UPL} provides the most efficient trade-off between accuracy and latency by a significant margin.
\acrshort{architecturename}\textit{-UPL} consistently outperforms the other architectures at each operating point, and fully defines the Pareto frontier of efficient solutions.
It is typical for \acrshort{architecturename}\textit{-UPL} to improve WER by over 8\% relative on LibriSpeech and 7\% on Voice Assistant of the nearest competitor for fixed UPL budgets.

These figures also show that providing \acrshort{architecturename}\textit{-UPL} models with more compute power naturally leads to better trade-offs between UPL and WER.
The higher throughput enables the models to apply lookahead attention to a greater degree without paying as high of a latency penalty since when the compute does become available (all DAG dependencies met), it executes quicker.
Furthermore, under this definition of latency, as long as the backlog of compute remaining is minimal on the final frame of each utterance, models have the ability to attend to future context as much as they want as long as the work can be burned down before the end of the utterance.
As the figures highlight, the  more compute throughput the model is given, the better it can take advantage of this property to improve its WER.

\section{Attention Dynamics}
\label{appendix:attn_dynamics}
\begin{figure}[H]
\centering
\subfigure{
    \centering
    \includegraphics[width=0.45\columnwidth]{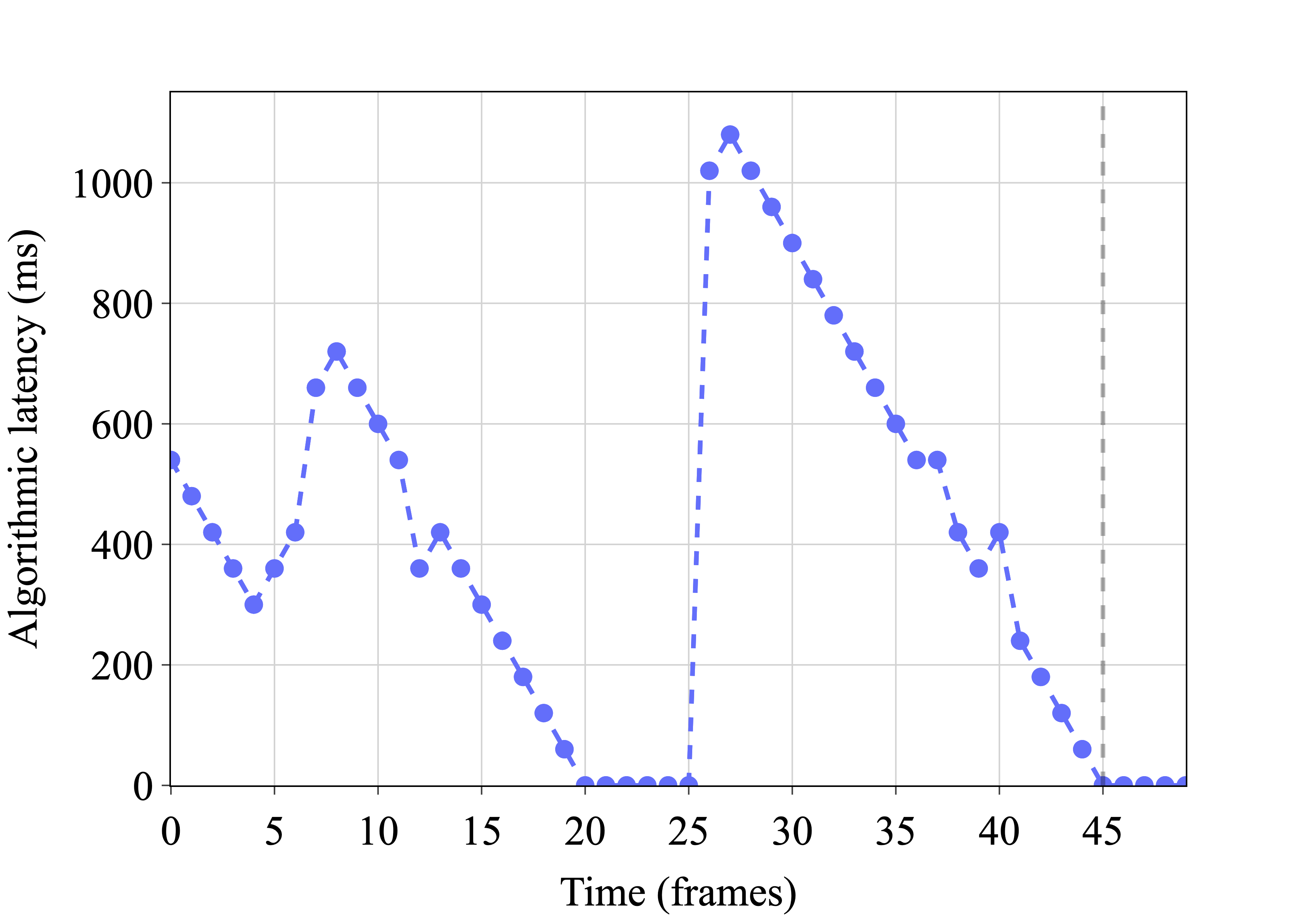}}%
\subfigure{
    \centering
    \includegraphics[width=0.45\columnwidth]{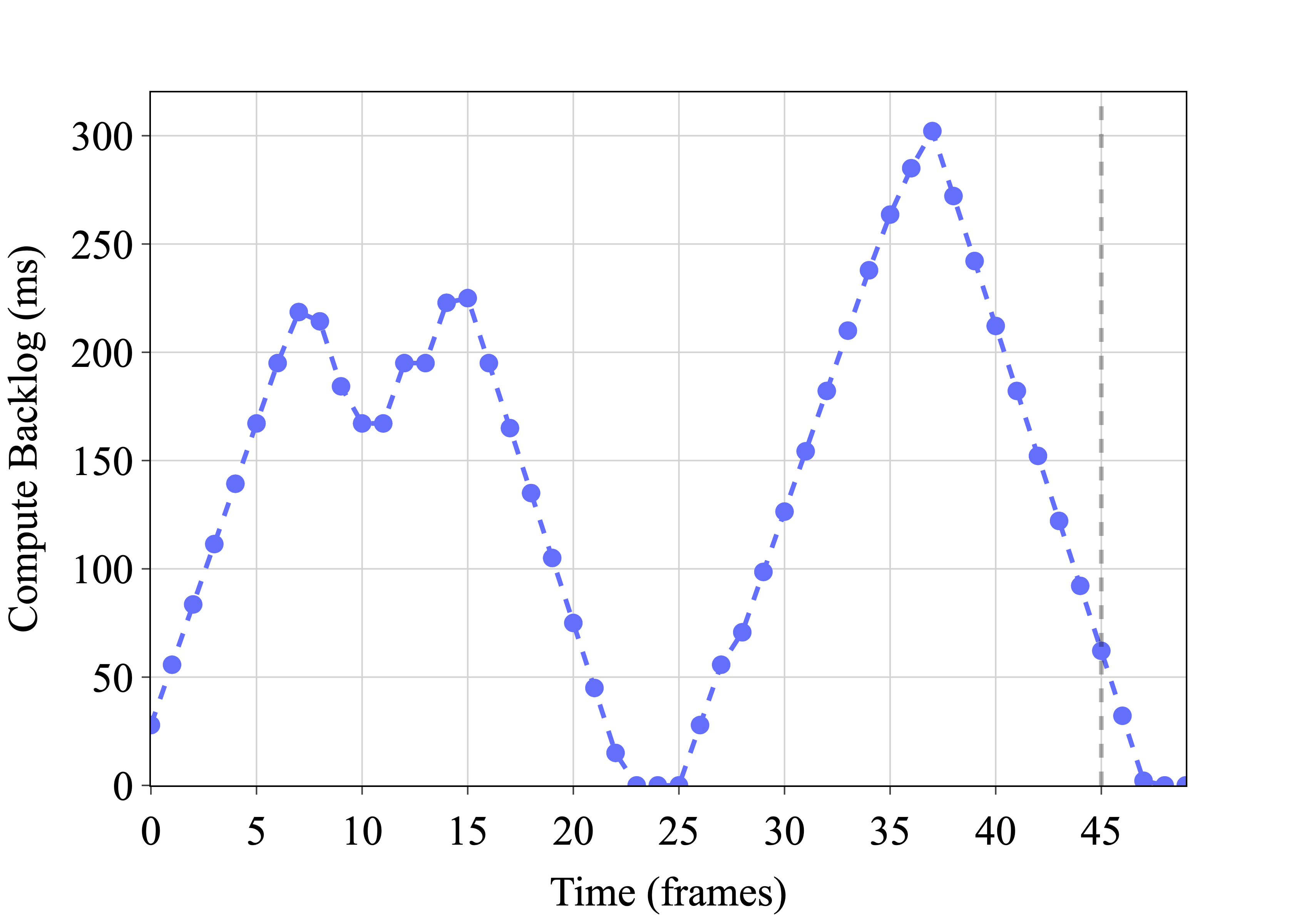}
}%
\caption{Voice Assistant Conformer \acrshort{architecturename} model frame-wise algorithmic latency (left) and the compute backlog present at each frame for an utterance (right). The utterance is 46 frames in length and a throughput of 2 is used for the compute backlog evaluations.}
\label{fig:va_al_upl_cpkt99_0}
\end{figure}

\begin{figure}[H]
\centering
\includegraphics[width=\columnwidth]{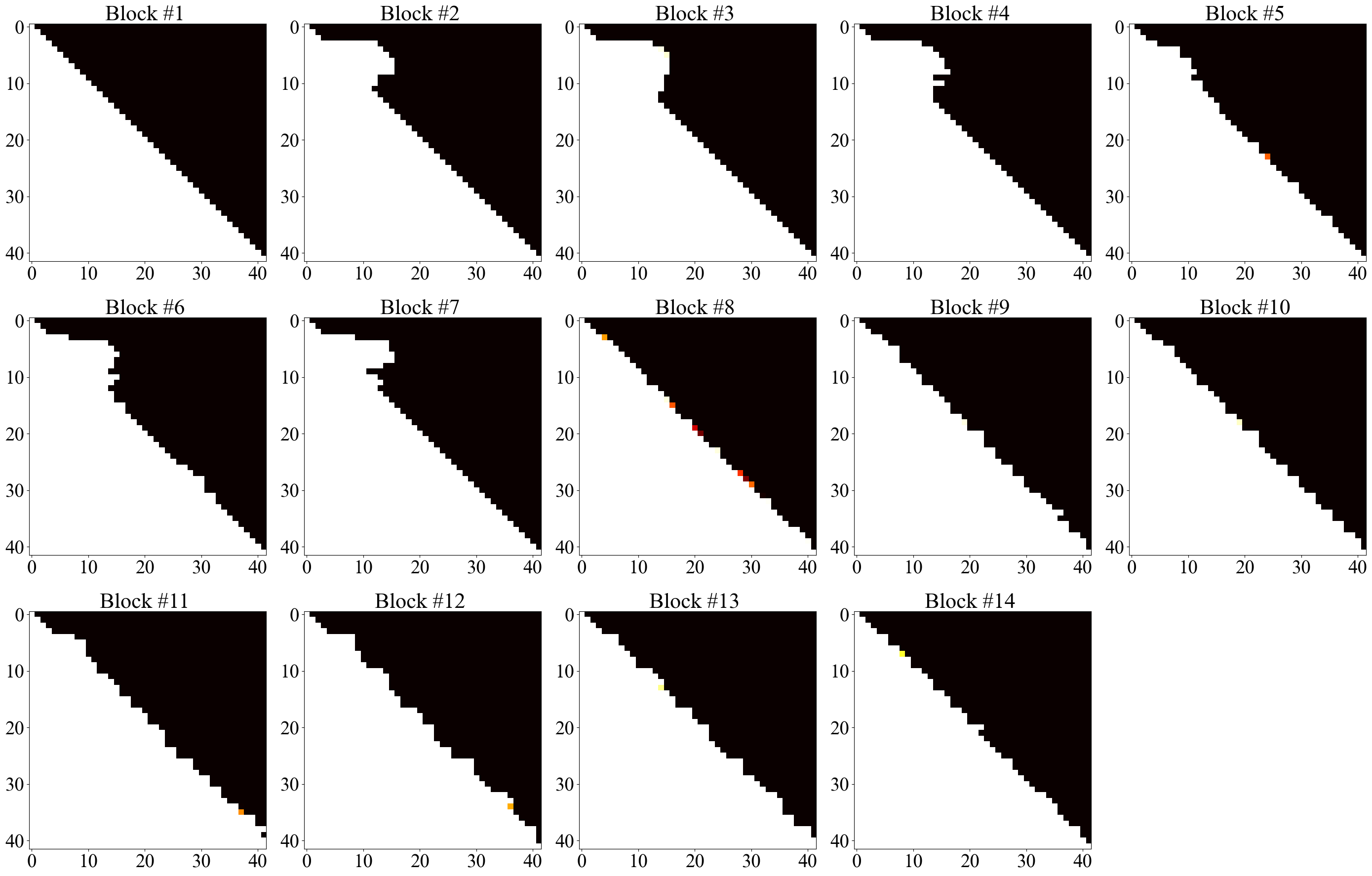}
\caption{LibriSpeech example utterance block-wise attention masks for Conformer \acrshort{architecturename}\textit{-Alg.Lat.} model.
One observes the distinctive stepwise patterns of \acrshort{architecturename}\textit{-Alg.Lat.} with large amounts of clustered lookahead at particular segments within the utterance.}
\label{fig:libri_attmaps}
\end{figure}

\begin{figure}[H]
\centering
\includegraphics[width=\columnwidth]{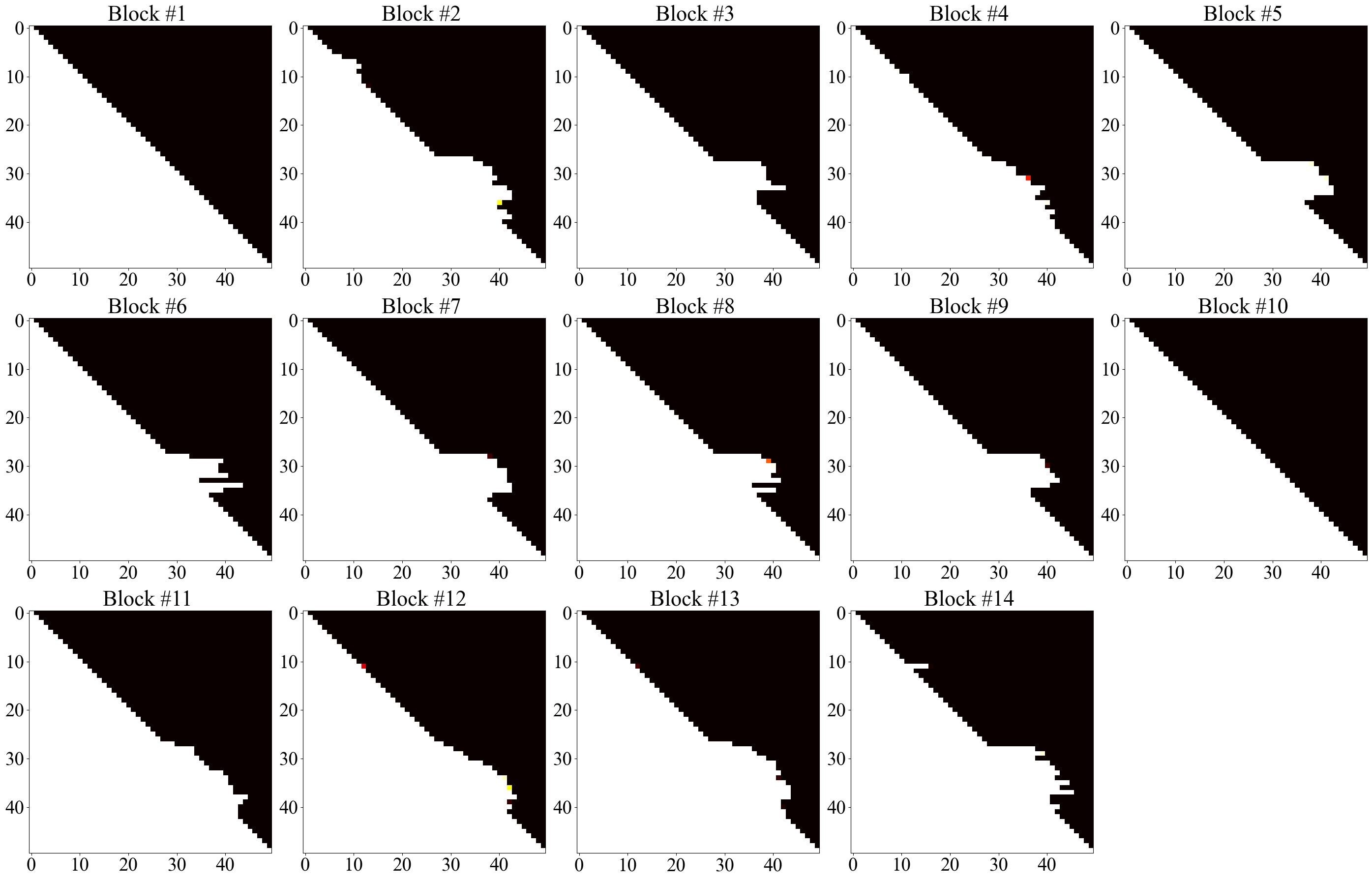}
\caption{Voice Assistant example utterance block-wise attention masks for Conformer \acrshort{architecturename}\textit{-Alg.Lat.} model.
One observes the distinctive stepwise patterns of \acrshort{architecturename}\textit{-Alg.Lat.} with large amounts of clustered lookahead at particular segments within the utterance.}
\label{fig:alexa_blockwise_attention}
\end{figure}

Plots are presented for frame-wise latency measures through an example utterance in Figure~\ref{fig:va_al_upl_cpkt99_0}, and full attention masks for all layers on both LibriSpeech and Voice Assistant data examples are also displayed in Figures~\ref{fig:libri_attmaps} and \ref{fig:alexa_blockwise_attention}.
Figure~\ref{fig:va_al_upl_cpkt99_0} shows that algorithmic latency can vary greatly over each frame of an utterance with spikes occurring in places where the model is less confident and desires further future context.
We see similar spikes for the accumulation and burn down of compute backlog throughout the utterance.
Note, however, that the rise and fall of the backlog is steadier throughout since progress completing nodes can more consistently occur as dependencies for intermediate layer nodes are met.
Also observe, that in this 46-frame example (2.76 seconds long), because there should be no forward attention past the end of the utterance, the algorithmic latency is rightfully 0 at the utterance boundary and beyond.
Likewise, because compute backlog defines the amount of work remaining and there is a non-zero backlog at the final frame, this utterance will observe a positive UPL.
Minimizing this backlog at the utterance boundary (frame index 45) is the objective of our UPL loss algorithm.

\section{Training Characteristics}
\label{appendix:training}
Throughout training, the temperature parameter $\tau$ is annealed from 1 to near 0, which transitions the soft DAG edges (and attention mask) into binary ones.
This appendix shows how the changes of $\tau$ throughout training impacts various components of the \acrshort{architecturename} architecture.

\begin{figure}[H]
\centering
\includegraphics[width=\columnwidth]{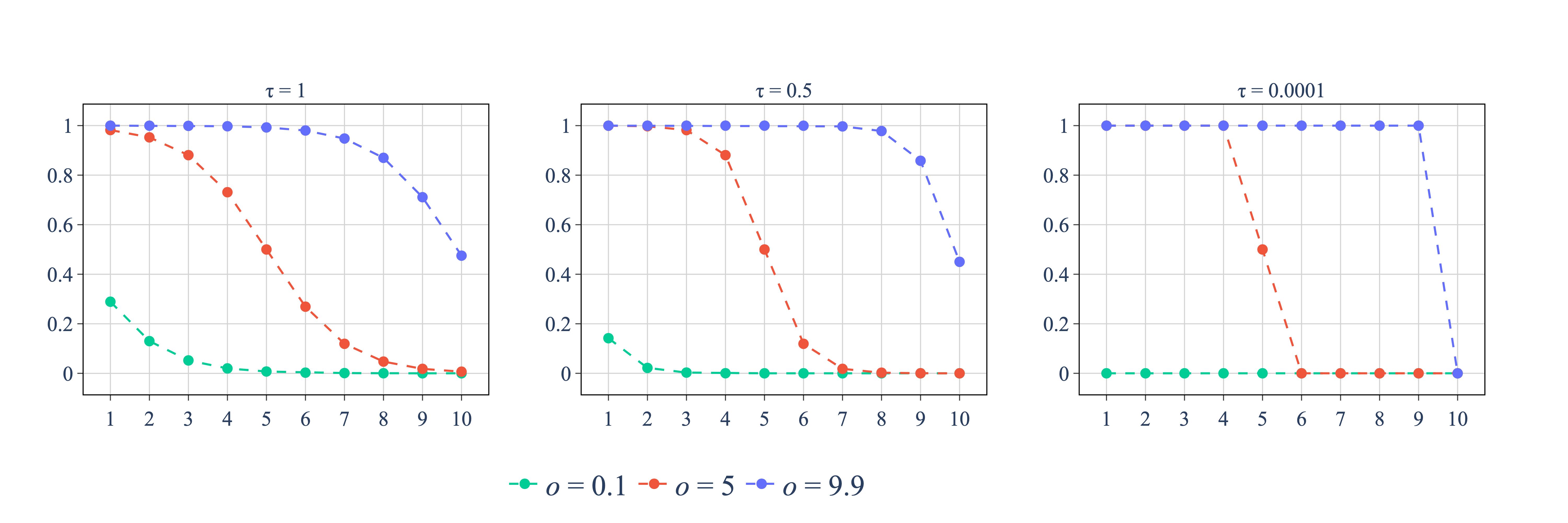}
\caption{Scheduler values for different predicted curve centers $o$ with a maximum span of $\forwardspan=10$ at various temperatures. These values mask the forward attention positions and represent the presence of a forward edge in the compute DAG. As temperature decreases, the edges become binarized.}
\label{fig:libri_attnew}
\end{figure}

Figure~\ref{fig:libri_attnew} shows the various future frame scheduler masking values at different temperatures.
Notice how the curve sharpens from a reverse ``S'' shape into a step function.

\begin{figure}[H]
\centering
\includegraphics[width=0.8\columnwidth]{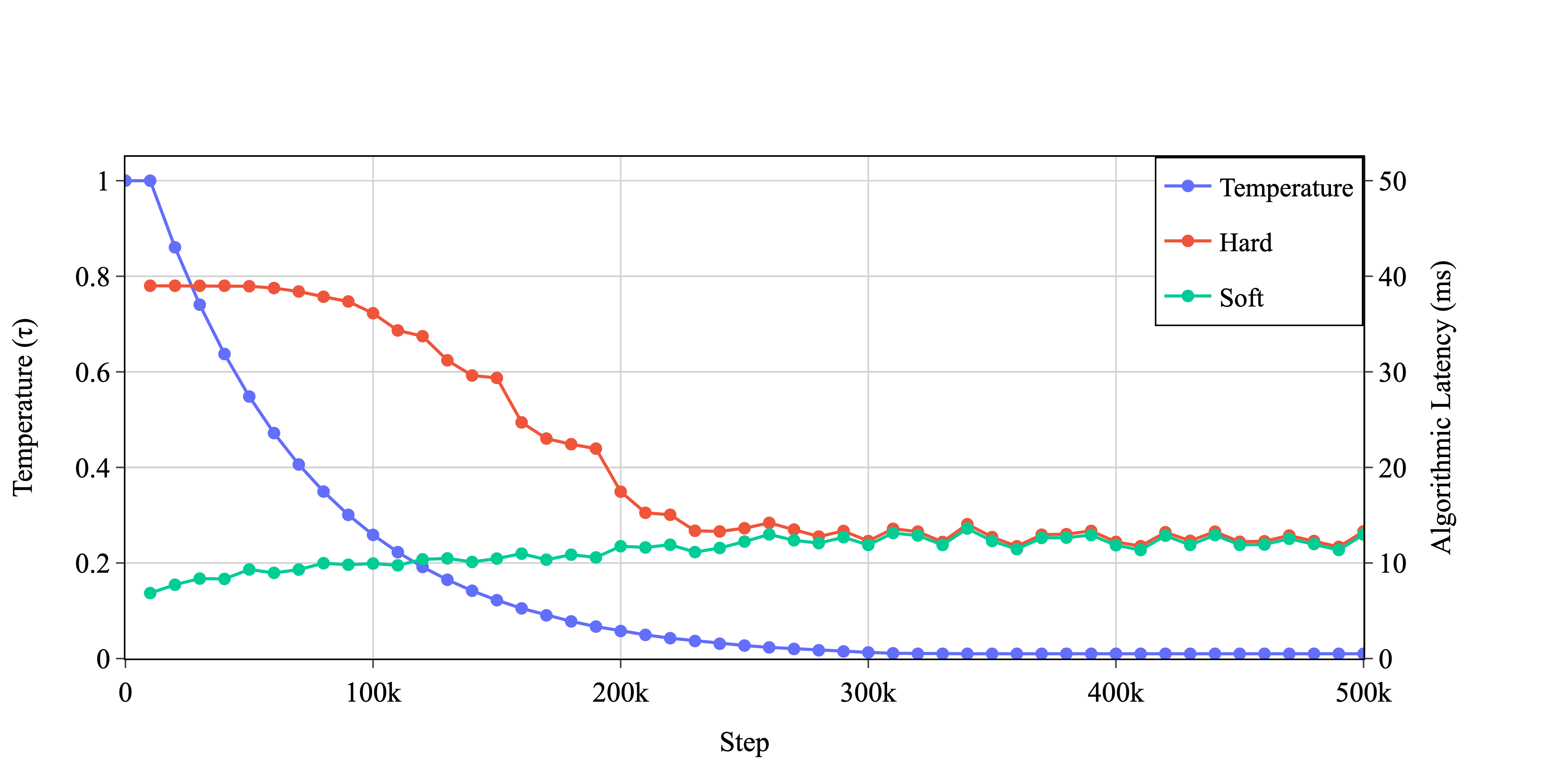}
\caption{Temperature, soft latency, and hard latency over training. Model checkpoints were evaluated every 10k steps with a Conformer \acrshort{architecturename}\textit{-Alg.Lat.} model.}
\label{fig:alexa_latency_temp_steps}
\end{figure}

During training, \acrshort{architecturename} has soft edges.
The loss function therefore computes a corresponding ``soft latency''. 
Meanwhile, the ``hard latency'', where all non-zero edge values are treated as 1, is used to evaluate the true run-time latency at test time.
Naturally these two measures are different at the start of training.
We show in Figure~\ref{fig:alexa_latency_temp_steps} however, that as temperature anneals over time, hard and soft latency measures smoothly converge to the ultimate final algorithmic latency, with the hard always an upper bound on soft.

\begin{figure}[H]
\centering
\includegraphics[width=\columnwidth]{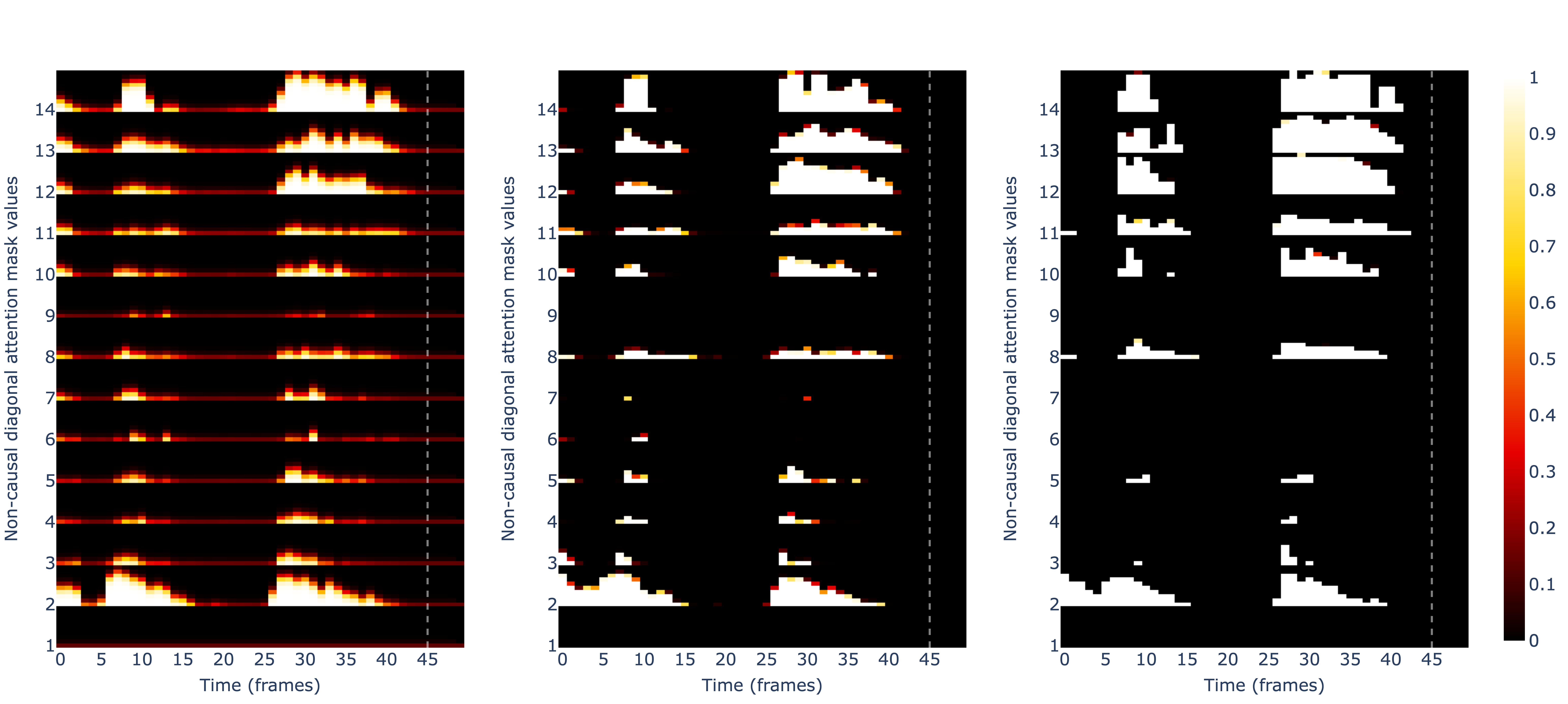}
\caption{Per-frame non-causal attention mask values of an utterance for a Conformer \acrshort{architecturename}\textit{-UPL} model. 
The mask values are derived from checkpoints evaluated at step 50k where temperature is 0.548 (left), at step 160k where temperature is 0.105 (center), and at step 500k, the final step, where temperature is 0.01 (right).
The utterance consists of a far-field user speaking ``[Voice Assistant], play Christmas music.'' 
The max forward span $\forwardspan$ of each layer is 10, and the figure shows only the 10 forward mask values right of the main diagonal of the full mask. The 10 forward mask values ascend from bottom to top.
Layers are labeled on the left vertical axes.
Throughout training, the attention masks shift from fractional to nearly binary.
Most of the attention structure is preserved but with definite differences as the model converges and adjusts itself to the annealing temperature.}
\label{fig:per_frame_non_causal_attention}
\end{figure}

\begin{figure}[H]
\centering
\includegraphics[width=\columnwidth]{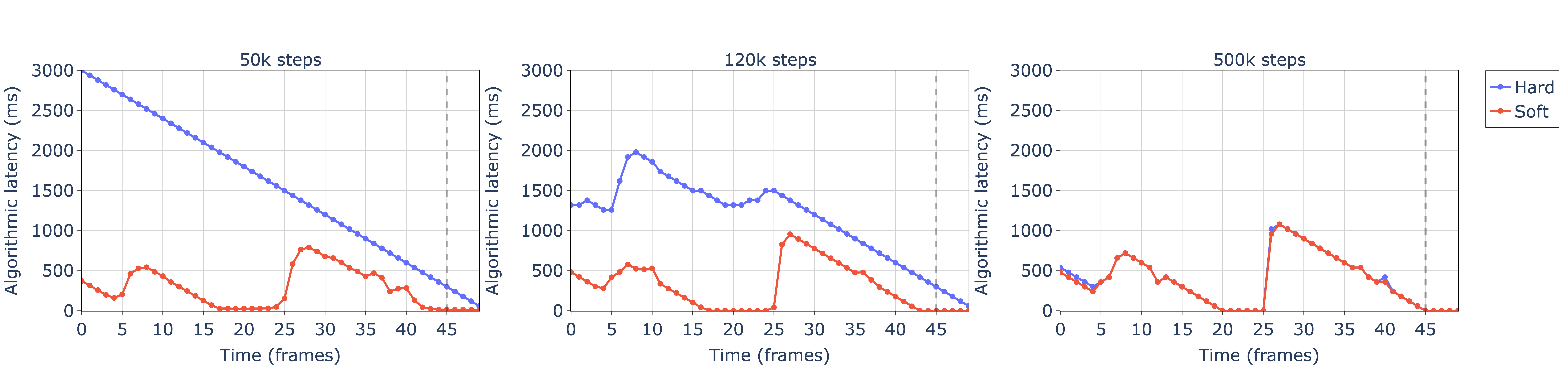}
\includegraphics[width=\columnwidth]{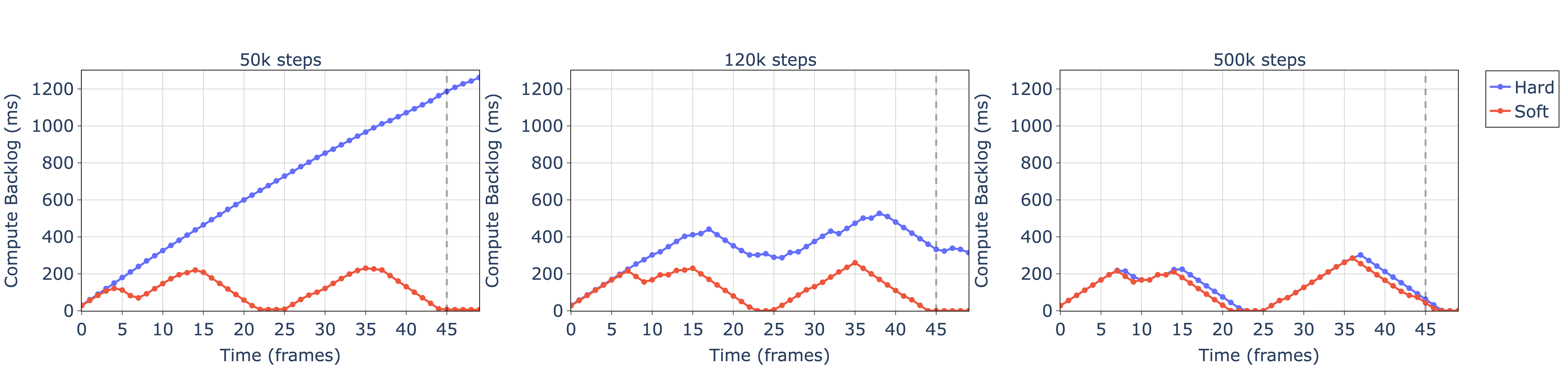}
\caption{Voice Assistant Conformer \acrshort{architecturename} model frame-wise algorithmic latency (top) and the compute backlog present
at each frame (bottom) for an utterance evaluated at three checkpoints during training: step 50k where temperature is 0.548 (left), step 160k where temperature is 0.105 (center), and step 500k the final step, where temperature is 0.01 (right). A throughput of 2 is used for the compute backlog evaluations.}
\label{fig:alexa_per_frame_plots}
\end{figure}

Figures~\ref{fig:per_frame_non_causal_attention} and \ref{fig:alexa_per_frame_plots} visualize per-frame attention, algorithmic latency, and compute backlog for an \acrshort{architecturename} model at three checkpoints during training for the same 46-frame utterance from the test set. 
Figure~\ref{fig:per_frame_non_causal_attention} shows the evolution of the lookahead attention masking values for the utterance and how they evolve to their final position.
One observes how the training begins with many fractional attention mask values but becomes more polarized (closer to 0 or 1) due to temperature annealing. 
For the last checkpoint, where the temperature $\tau$ is 0.01, attention masking is nearly binary.

Next, we visualize the characteristics of per-frame soft and hard latency throughout training in Figure~\ref{fig:alexa_per_frame_plots}.
First, observe that although per-frame soft and hard algorithmic latency patterns differ greatly at earlier epochs, they converge to nearly the same values at the final checkpoint as expected.
See also in the first checkpoint that hard algorithmic latency is at its maximum, matching what one would expect from a full-context attention model, and similarly at this checkpoint, the backlog gradually builds over time as more work than can be executed in real-time is added frame over frame.

Next, comparing frame-wise latency between checkpoints, we see soft algorithmic latency keeping a similar shape, but the amplitude increases slightly over the course of training.
This suggests that earlier in training, the model can learn to keep important attention weights small, yet non-zero, to minimize the soft penalty term. 
As temperature decreases however, these non-zero weights become increasingly polarized where the now stronger weighted forward connections cause higher algorithmic latency.
This matches Figure~\ref{fig:alexa_latency_temp_steps} findings.

Finally, we discuss the per-frame compute backlog. 
Similar to the trends we saw with algorithmic latency, we see soft and hard latency converge over the epochs.
Soft backlog maintains the same shape but increases in amplitude.
The hard backlog starts at a maximum and then decreases to converge with the soft backlog calculation. 
Taking a closer examination of the hard backlog in the first checkpoint, one notices the backlog continues to grow frame by frame due to the large hard algorithmic latency. 
Close examination reveals that at every 10 frames, there is a slight decrease in slope. 
This phenomenon occurs because after 10 frames, the nodes for the first layer start to become available, since the max per-layer right context is 10 frames.
After 20 frames, the second layer starts to become available, and so on every 10 frames. 
If the utterance were long enough, we would see that the hard compute backlog would reach its maximum value at 140 frames (14 layers $\times$ 10 lookahead). 

\section{Training Configuration Specifics}
\label{append:libri-setting}
All the models presented in this paper on LibriSpeech and Voice Assistant data are trained on 6 $\times$ 8 NVIDIA Tesla V100 GPUs with a per-core bucket batch size of [32, 16, 12, 4] according to different utterance lengths. For our LibriSpeech \acrshort{architecturename} models, we initially set the temperature for schedulers as $\tempparam = 1$ and anneal it to $1\mathrm{e}{-4}$ with an exponential decay rate of $0.99996$ beginning at epoch 40.
Voice Assistant uses a longer decay annealing over 320k steps to a temperature of 0.01.
We also list the detailed encoder configuration of the Conformer model in Table~\ref{tab:model-size}. The T-T architecture we use in the experiments has the same hyperparameters except those of the convolutions models which do not apply.
The scheduler networks consist of a total of 0.2M parameters which are considerably lightweight compared to our main Transformer-based encoder.
\vspace{5mm}
\begin{table}[H]
\centering
\caption{Encoder Model Hyperparameters for Conformer.}
\begin{tabular}{cc}
\hline
\multicolumn{2}{c}{Conformer Encoder Setup} \\ \hline
Num of Params (M) & 27.5  \\ 
Encoder Layers & 14 \\
Encoder Dim & 256            \\ 
Num of Attention Heads                          & 4            \\
Feed-Forward Dim                               & 1024           \\
Point-wise-1 Conv Kernel Size & (512, 256, 1, 1) \\
Depth-wise Conv Kernel Size & (32, 1, 1, 1) \\
Point-wise-2 Conv Kernel Size & (256, 512, 1, 1) \\
Output Size                            & 512            \\ \hline
\end{tabular}
\label{tab:model-size}
\end{table}
\vspace{5mm}




\section{Voice Assistant Result Tables}
\label{appndx:tables}
For completeness, we also construct a table for Voice Assistant data as done with LibriSpeech for both Conformer and T-T baselines and \acrshort{architecturename} models. We provide model label names for ease of reference.

\begin{table}[htb!]
\centering
\begin{tabular}{lllrrrrr}
\toprule
\thead{Label} & \thead{Attention \\ (right context \\ frames)} & \thead{Regularization \\ Term} & \thead{Reg. \\ weight} & \thead{Throughput} & \thead{WERR} & \thead{Alg. \\ Latency \\ (ms)} & \thead{UPL \\ (ms)} \\
\midrule
Baseline & Causal & N/A & N/A & 2.0 & 0.0 & 0.0 & 0.0 \\
Layer1 & Layerwise (1) & N/A & N/A & 2.0 & -9.2 & 797.3 & 224.2 \\
Layer2 & Layerwise (2) & N/A & N/A & 2.0 & -11.9 & 1399.7 & 446.7 \\
Layer10 & Layerwise (10) & N/A & N/A & 2.0 & -21.1 & 2323.0 & 1160.8 \\
Chunked5 & Chunked (5) & N/A & N/A & 2.0 & -3.9 & 122.1 & 82.8 \\
Chunked10 & Chunked (10) & N/A & N/A & 2.0 & -8.2 & 277.5 & 194.7 \\
Chunked20 & Chunked (20) & N/A & N/A & 2.0 & -11.2 & 583.2 & 402.3 \\
Chunked30 & Chunked (30) & N/A & N/A & 2.0 & -12.0 & 871.4 & 681.6 \\
L1-A & \acrshort{architecturename} (10) & L1 & 1.00e-04 & 2.0 & -16.8 & 1519.0 & 772.5 \\
L1-B & \acrshort{architecturename} (10) & L1 & 2.50e-04 & 2.0 & -14.4 & 1177.9 & 606.8 \\
L1-C & \acrshort{architecturename} (10) & L1 & 5.00e-04 & 2.0 & -12.9 & 825.2 & 361.4 \\
L1-D & \acrshort{architecturename} (10) & L1 & 7.50e-04 & 2.0 & -11.2 & 619.4 & 239.8 \\
L1-E & \acrshort{architecturename} (10) & L1 & 1.00e-03 & 2.0 & -9.4 & 587.5 & 236.7 \\
L1-F & \acrshort{architecturename} (10) & L1 & 1.25e-03 & 2.0 & -8.7 & 412.7 & 150.0 \\
L1-G & \acrshort{architecturename} (10) & L1 & 1.50e-03 & 2.0 & -7.0 & 266.0 & 33.4 \\
L1-H & \acrshort{architecturename} (10) & L1 & 1.75e-03 & 2.0 & -5.8 & 242.1 & 21.9 \\
L1-I & \acrshort{architecturename} (10) & L1 & 2.00e-03 & 2.0 & -6.6 & 188.1 & 11.2 \\
L1-J & \acrshort{architecturename} (10) & L1 & 2.50e-03 & 2.0 & -5.2 & 206.8 & 22.4 \\
AL-A & \acrshort{architecturename} (10) & Alg. Latency & 1.25e-03 & 2.0 & -17.1 & 1007.9 & 712.5 \\
AL-B & \acrshort{architecturename} (10) & Alg. Latency & 2.50e-03 & 2.0 & -13.9 & 884.8 & 628.0 \\
AL-C & \acrshort{architecturename} (10) & Alg. Latency & 3.75e-03 & 2.0 & -14.3 & 657.1 & 467.4 \\
AL-D & \acrshort{architecturename} (10) & Alg. Latency & 5.00e-03 & 2.0 & -11.7 & 503.4 & 344.4 \\
AL-E & \acrshort{architecturename} (10) & Alg. Latency & 7.50e-03 & 2.0 & -9.5 & 256.7 & 46.6 \\
AL-F & \acrshort{architecturename} (10) & Alg. Latency & 1.00e-02 & 2.0 & -8.9 & 176.3 & 12.0 \\
AL-G & \acrshort{architecturename} (10) & Alg. Latency & 1.50e-02 & 2.0 & -6.5 & 88.2 & 3.1 \\
AL-H & \acrshort{architecturename} (10) & Alg. Latency & 1.75e-02 & 2.0 & -4.8 & 81.8 & 2.5 \\
CB01.5-A & \acrshort{architecturename} (10) & UPL & 8.75e-05 & 1.5 & -14.8 & 888.1 & 264.5 \\
CB01.5-B & \acrshort{architecturename} (10) & UPL & 1.30e-04 & 1.5 & -14.9 & 725.2 & 154.4 \\
CB01.5-C & \acrshort{architecturename} (10) & UPL & 1.93e-04 & 1.5 & -14.0 & 611.8 & 107.6 \\
CB01.5-D & \acrshort{architecturename} (10) & UPL & 4.38e-04 & 1.5 & -10.0 & 526.5 & 58.5 \\
CB01.5-E & \acrshort{architecturename} (10) & UPL & 9.77e-04 & 1.5 & -9.9 & 455.9 & 22.8 \\
CB2-A & \acrshort{architecturename} (10) & UPL & 1.17e-04 & 2.0 & -16.9 & 814.9 & 174.4 \\
CB2-B & \acrshort{architecturename} (10) & UPL & 1.73e-04 & 2.0 & -17.9 & 796.0 & 125.6 \\
CB2-C & \acrshort{architecturename} (10) & UPL & 2.57e-04 & 2.0 & -14.3 & 642.2 & 72.0 \\
CB2-D & \acrshort{architecturename} (10) & UPL & 5.83e-04 & 2.0 & -13.4 & 585.2 & 47.6 \\
CB2-E & \acrshort{architecturename} (10) & UPL & 1.30e-03 & 2.0 & -12.4 & 482.2 & 17.1 \\
CB3-A & \acrshort{architecturename} (10) & UPL & 1.75e-04 & 3.0 & -17.4 & 890.0 & 108.4 \\
CB3-B & \acrshort{architecturename} (10) & UPL & 2.59e-04 & 3.0 & -17.1 & 722.6 & 51.9 \\
CB3-C & \acrshort{architecturename} (10) & UPL & 3.85e-04 & 3.0 & -14.7 & 673.6 & 48.2 \\
CB3-D & \acrshort{architecturename} (10) & UPL & 8.75e-04 & 3.0 & -13.8 & 594.5 & 24.2 \\
CB3-E & \acrshort{architecturename} (10) & UPL & 1.95e-03 & 3.0 & -12.3 & 550.6 & 12.2 \\
\bottomrule
\end{tabular}
\caption{Voice Assistant Conformer Experimental Results.}
\label{table_alexa_conf}
\end{table}


\begin{table}[htb!]
\centering
\begin{tabular}{lllrrrrr}
\toprule
\thead{Label} & \thead{Attention \\ (right context \\ frames)} & \thead{Regularization \\ Term} & \thead{Reg. \\ weight} & \thead{Throughput} & \thead{WERR} & \thead{Alg. \\ Latency \\ (ms)} & \thead{UPL \\ (ms)} \\
\midrule
Baseline & Causal & N/A & N/A & 2.0 & 0.0 & 0.0 & 0.0 \\
Layer10 & Layerwise (10) & N/A & N/A & 2.0 & -18.1 & 2322.7 & 1160.9 \\
Chunked10 & Chunked (10) & N/A & N/A & 2.0 & -5.5 & 277.5 & 194.7 \\
Chunked20 & Chunked (20) & N/A & N/A & 2.0 & -9.5 & 583.3 & 402.4 \\
L1 & \acrshort{architecturename} (10) & L1 & 5.00e-04 & 2.0 & -15.6 & 967.6 & 310.5 \\
AL & \acrshort{architecturename} (10) & Alg. Latency & 5.00e-03 & 2.0 & -16.4 & 563.8 & 292.0 \\
\bottomrule
\end{tabular}
\caption{Voice Assistant Transformer-Transducer Experimental Results.}
\label{table_alexa_tt}
\end{table}

\end{document}